\documentclass{elsart}
\usepackage{amsmath,amssymb}
\usepackage{cite}
\usepackage{graphicx}
\usepackage{subfigure}
\usepackage[bf,footnotesize]{caption}
\usepackage{hyperref}
\usepackage{appendix}

\graphicspath{{plots/}}
\def \3{\ss }

\newcommand{\tr}{\operatorname{Tr}}
\newcommand{\re}{\operatorname{Re}}

\newcommand{\mps}{m_\mathrm{PS}}
\newcommand{\fps}{f_\mathrm{PS}}
\newcommand{\mpcac}{m_\mathrm{PCAC}}

\newcommand{\tauint}{\tau_\mathrm{int}}
\newcounter{bla}

\begin{document}
\begin{frontmatter}

\title{Light Meson Physics from Maximally Twisted Mass Lattice QCD}

\begin{center}
  \includegraphics[draft=false]{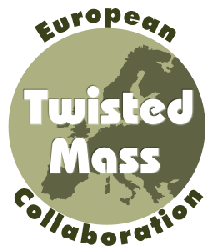}
\end{center}

\author{ETM Collaboration},
\author[a]{R. Baron},
\author[a]{Ph. Boucaud},
\author[b]{P. Dimopoulos},
\author[c]{F. Farchioni},
\author[b]{R. Frezzotti},
\author[d]{V. Gimenez},
\author[e]{G. Herdoiza},
\author[e]{K. Jansen},
\author[f]{V. Lubicz},
\author[g]{C. Michael},
\author[c]{G. M\"unster},
\author[b]{D. Palao},
\author[b]{G.C. Rossi},
\author[k]{L. Scorzato},
\author[h]{A. Shindler},
\author[f]{S. Simula},
\author[c]{T. Sudmann},
\author[i]{C. Urbach\thanksref{author}},
\author[j]{U. Wenger}

\thanks[author]{Corresponding author}

\address[a]{ Laboratoire de Physique Th\'eorique (B\^at.~210), Universit\'e
      de Paris XI, Centre d'Orsay, 91405 Orsay-Cedex, France}
\address[b]{Dip. di Fisica, Universit{\`a} di Roma Tor Vergata and INFN,
      Sez. di Tor Vergata, Via della Ricerca Scientifica, I-00133 Roma, Italy}
\address[c]{Universit\"at M\"unster, Institut f\"ur Theoretische Physik,
      Wilhelm-Klemm-Strasse 9, D-48149 M\"unster, Germany}
\address[d]{Dep. de F\'{\i}sica Te\`{o}rica and IFIC, Universitat de Val\`encia-CSIC,
  Dr. Moliner 50, E-46100 Burjassot, Spain}
\address[e]{NIC, DESY, Zeuthen, Platanenallee 6, D-15738 Zeuthen, Germany}
\address[f]{Dip. di Fisica, Universit{\`a} di Roma Tre and INFN, Sez. di
      Roma III, Via della Vasca Navale 84, I-00146 Roma, Italy}
\address[g]{Theoretical Physics Division, Dept. of Mathematical Sciences,
      University of Liverpool, Liverpool L69 7ZL, UK}
\address[k]{ECT*, strada delle tabarelle 286, 38100Trento, Italy}
\address[h]{Instituto de Fisica Teorica, Universidad Autonoma de
  Madrid, Facultad de Ciencias C-XI, E-28049 Cantoblanco, Madrid, Spain}
\address[i]{Helmholtz-Institut f{\"u}r Strahlen- und Kernphysik
  (Theorie) and Bethe Center for Theoretical Physics, Universit{\"a}t
  Bonn, 53115 Bonn, Germany}
\address[j]{Institute for Theoretical Physics, University of Bern,
  Sidlerstr. 5, CH-3012 Bern, Switzerland}

\clearpage
\begin{abstract}

\noindent We present a comprehensive investigation of light meson physics using
maximally twisted mass fermions for $N_f=2$ mass-degenerate quark
flavours.  By employing four values of the lattice spacing, spatial lattice
extents ranging from 2.0\,fm to 2.5\,fm and pseudo scalar masses in
the range $280\lesssim m_\mathrm{PS}\lesssim 650$\,MeV we control the
major systematic effects of our calculation. This enables us to
confront our data with chiral perturbation theory and extract low
energy constants of the effective chiral Lagrangian and derived
quantities, such as the light quark mass, with high precision.

\begin{flushleft}
Preprint-No: HISKP-TH-09-37, DESY 09-187, SFB/CPP-09-115, MS-TP-09-27,
ROM2F/2009/24, IFT-UAM/CSIC-09-58
\end{flushleft}


\end{abstract}

\end{frontmatter}

\section{Introduction}


In recent years, the non-perturbative description of QCD on the
lattice has made significant progress in tackling systematic effects
present in the determination of several important physical quantities
(see e.g. \cite{Jansen:2008vs,Scholz:2009yz} for recent reviews).  Simulations
containing the dynamics of the light-quark flavours in the sea, as
well as those due to the strange quark and recently also to the charm,
using pseudo scalar masses below $300$\,MeV, spatial lattice extents $L \geq
2$\,fm and lattice spacings smaller than $0.1$\,fm are presently
being performed by several lattice groups. Such simulations will
eventually allow for an extrapolation of the lattice data to the
continuum limit and to the physical point while keeping also the
finite volume effects under control.

Adding a twisted mass term \cite{Frezzotti:2000nk} to the 
Wilson-Dirac operator and tuning the theory to 
maximal twist \cite{Frezzotti:2003ni,Frezzotti:2004wz}
has by now proved to be a practical and successful tool for performing
$\mathcal{O}(a)$-improved lattice QCD simulations, see e.g.
refs.~\cite{Boucaud:2007uk,Blossier:2007vv,Cichy:2008gk,Boucaud:2008xu,Alexandrou:2008tn,Jansen:2008wv,Shindler:2007vp,Blossier:2009bx,Jansen:2009hr,McNeile:2009mx,Blossier:2009hg,Jansen:2009tt}.
A detailed study of the continuum-limit scaling in the quenched approximation 
\cite{Jansen:2003ir,Jansen:2005gf,Jansen:2005kk,Abdel-Rehim:2005gz} revealed 
not only that with an appropriate tuning procedure to maximal twist 
lattice artefacts follow the expected \cite{Frezzotti:2003ni} 
$\mathcal{O}(a^2)$ scaling behaviour, but also that the 
remaining $\mathcal{O}(a^2)$ effects are small, in agreement
with the conclusions drawn in ref.~\cite{Frezzotti:2005gi}. The only
exception seen so far is the neutral pseudo scalar mass
\cite{Jansen:2005cg} which shows significant $\mathcal{O}(a^2)$ effects.
We will discuss this issue and its interpretation \cite{Frezzotti:2007qv,Dimopoulos:2009qv} further
in sec.~3.4.1 below.

In this paper, we shall present a study of the continuum limit scaling
of lattice QCD with maximally twisted mass fermions for the case of
dynamical fermions with $N_f=2$ mass degenerate quarks demonstrating that 
also in this setup the $\mathcal{O}(a^2)$ effects are small 
as has already been discussed 
in refs.~\cite{Urbach:2007rt,Dimopoulos:2007qy,Dimopoulos:2008sy}.
We use data for the charged pseudo scalar meson mass
$m_\mathrm{PS}$ and the decay constant $f_\mathrm{PS}$ 
computed in a range of 
$280\, \mathrm{MeV}\lesssim m_\mathrm{PS} \lesssim 650\, \mathrm{MeV}$. 
Our analysis is concentrated on 
two values of the inverse bare coupling $\beta$, 
$\beta=3.9$ and $\beta=4.05$ corresponding to values of the lattice 
spacing of $a=0.079\, \mathrm{fm}$ and $a=0.063\, \mathrm{fm}$, respectively. 
We complement our data set by adding an additional coarser lattice 
spacing of $a=0.100\, \mathrm{fm}$ corresponding to $\beta=3.8$ and 
a finer lattice spacing 
of $a=0.051\, \mathrm{fm}$, corresponding to $\beta=4.2$.
Our new results at 
$\beta=4.2$, computed at two values of the pseudo scalar mass, confirm 
the smallness of lattice artefacts in the quantities considered here.

The smallness of lattice artefacts, the range of masses covered, 
the usage of several lattice volumes  
and the precision with which $m_\mathrm{PS}$ and $f_\mathrm{PS}$
can be obtained using maximally twisted mass fermions
allow us to confront the numerical data with chiral 
perturbation theory and to eventually extract accurate values for a number
of low energy constants, in particular $\bar{\ell}_3=3.50(31)$ and
$\bar{\ell}_4=4.66(33)$, as will be discussed below. The main physical
results we obtain from this analysis are the light quark mass
$m_{u,d}^{\overline{\mathrm{MS}}}(\mu=2\, \mathrm{GeV})=3.54(26)\,
\mathrm{MeV}$, the pseudo scalar decay constant in the chiral limit 
$f_0=122(1)\, \mathrm{MeV}$, the scalar condensate
$[\Sigma^{\overline{\mathrm{MS}}}(\mu=2\, \mathrm{GeV})]^{1/3}=270(7)\,
\mathrm{MeV}$ and $f_\pi/f_0=1.0755(94)$. The errors are statistical
and systematical errors summed in quadrature. In case of asymmetric
errors we use conservatively the maximum for the sum. The detailed
error budget can be found in table~\ref{tab:results} below.

A special emphasis of this paper is the investigation of systematic
errors on the results for the low energy constants from the above
mentioned chiral fits.  To estimate the systematic errors, we perform
different kind of fits where we take lattice spacing, finite size and
next-to-next-leading order (NNLO) corrections into account.  Finally,
we address the question of isospin violations by comparing the neutral
to charged pseudo scalar masses as well as discussing the isospin
splittings for other physical quantities.

The paper is organised as follows: after describing the lattice set-up
in the remainder of this section we shall present the main results of
this paper in section~\ref{sec:mainres}. The following
section~\ref{sec:ens} introduces the simulation set-up as well as the
main ingredients entering in the continuum-limit scaling analysis of
light meson observables. In section~\ref{sec:comb} we present the
results of the combined chiral perturbation theory fits and
section~\ref{sec:summ} gives a summary of our results. The details of
the analyses and data tables are collected in the appendices.

\subsection{Lattice Action}

The twisted mass action, the tuning procedure to maximal twist
and the analysis techniques as employed by our European Twisted Mass
Collaboration (ETMC) have been discussed extensively in 
refs.~\cite{Boucaud:2007uk,Boucaud:2008xu}. 
We therefore only briefly recapitulate the essential ingredients 
of our set-up.

In the gauge sector we employ the tree-level Symanzik
improved gauge action (tlSym) \cite{Weisz:1982zw}, viz.
\[
S_g = \frac{\beta}{3}\sum_x\left(  b_0\sum_{\substack{
    \mu,\nu=1\\1\leq\mu<\nu}}^4\{1-\re\tr(U^{1\times1}_{x,\mu,\nu})\}\Bigr. 
\Bigl.\ +\ 
b_1\sum_{\substack{\mu,\nu=1\\\mu\neq\nu}}^4\{1
-\re\tr(U^{1\times2}_{x,\mu,\nu})\}\right)\,  ,
\]
with the bare inverse gauge coupling $\beta=6/g_0^2$, $b_1=-1/12$ and
$b_0=1-8b_1$. The fermionic action for two flavours of 
twisted, mass degenerate quarks in the so called twisted
basis~\cite{Frezzotti:2000nk,Frezzotti:2003ni} reads
\begin{equation}
  \label{eq:sf}
  S_\mathrm{tm}\ =\ a^4\sum_x\left\{ \bar\chi(x)\left[ D[U] + m_0 +
      i\mu_q\gamma_5\tau^3\right]\chi(x)\right\}\, ,
\end{equation}
where $m_0$ is the untwisted bare quark mass, $\mu_q$ is the bare
twisted quark mass, $\tau^3$ is the third Pauli matrix acting in
flavour space and
\[
D[U] = \frac{1}{2}\left[\gamma_\mu\left(\nabla_\mu +
    \nabla^*_\mu\right) -a\nabla^*_\mu\nabla_\mu \right]
\]
is the mass-less Wilson-Dirac operator. $\nabla_\mu$ and $\nabla_\mu^*$
are the forward and backward gauge covariant difference operators,
respectively. Twisted mass fermions are said to be at \emph{maximal
 twist} if the bare untwisted quark mass $m_0$ is tuned to its critical
value $m_\mathrm{crit}$, the situation we shall be interested in. For 
convenience we define the hopping parameter $\kappa=1/(8+2am_0)$. Note
that we shall use the twisted basis throughout this paper.

Maximally twisted mass fermions provide important advantages over 
Wilson's originally proposed formulation of lattice QCD:
the spectrum of $Q^\dagger Q$ with 
$Q=\gamma_5(D[U]+m_0+i\mu_q\gamma_5)$ is bounded
from below, which was the original reason to consider twisted mass
fermions \cite{Frezzotti:2000nk}. 
At maximal twist, the twisted quark mass $\mu_q$ is related 
directly to the physical
quark mass and renormalises multiplicatively only. Many mixings under
renormalisation are expected to be simplified
\cite{Frezzotti:2003ni,Frezzotti:2004wz}. And -- most
importantly -- as was first shown in ref.~\cite{Frezzotti:2003ni},
physical observables are automatically $\mathcal{O}(a)$ improved
without the need to determine any operator-specific improvement
coefficients. Another feature of maximally twisted mass fermions is
that the pseudo scalar decay constant $f_\mathrm{PS}$ does not need
any renormalisation which allows for a very precise determination of
this quantity.

The main drawback of maximally twisted mass fermions is that both parity
and flavour symmetry are broken explicitly at non-zero values of the
lattice spacing.  However, it turns out that this is presumably only
relevant for the mass of the neutral pseudo scalar meson (and
kinematically related quantities). We shall discuss this issue in more
detail later on. Note that in the following we use the notation
$\mps\equiv\mps^\pm$ for the charged pion mass, while the neutral one
will be denoted by $\mps^0$.

Tuning to maximal twist is achieved with 
the general prescription \cite{Farchioni:2004ma,Farchioni:2004fs} 
to choose a parity odd 
operator $O$ and determine $am_\mathrm{crit}$ such that $O$ has
vanishing expectation value at fixed physical situation for all
lattice spacings, e.g. by fixing $m_\mathrm{PS}$ in physical units.             
One appropriate quantity to tune to zero is the PCAC quark
mass, defined as
\begin{equation}
  \label{eq:mpcac}
  m_\mathrm{PCAC} =
  \frac{\sum_\mathbf{x}\langle\partial_0A^a_0(\mathbf{x},t)P^a(0)\rangle}%
  {2\sum_\mathbf{x}\langle P^a(\mathbf{x},t)P^a(0)\rangle} \, ,\qquad\quad a=1,2\, ,
\end{equation}
where $A_\mu^a$ and $P^a$ are the axial vector current and the pseudo
scalar density in the twisted basis, respectively,
\[
A_\mu^a(x) = \bar\chi(x)\gamma_\mu\gamma_5\frac{\tau^a}{2}\chi(x)\,
,\qquad\qquad P^a(x) = \bar\chi(x)\gamma_5\frac{\tau^a}{2}\chi(x)\, .
\]
Once rotated to the physical basis the numerator of the r.h.s of
eq.~(\ref{eq:mpcac}) represents the vacuum expectation value of a
parity odd operator. Maximal twist is then achieved by demanding that
the PCAC quark mass of  
eq.~(\ref{eq:mpcac}) vanishes. A discussion about the precise conditions
we employ in our simulations is given in ref.~\cite{Boucaud:2008xu} and 
summarised in section~\ref{sec:ens}.

\section{Main Results}
\label{sec:mainres}

In this section we present the main results of this paper. All the
remaining sections are devoted to the discussion of the
details leading to these results and to the estimation of systematic
uncertainties.

The ETM collaboration has gathered data for the pseudo scalar decay
constant $\fps$ and mass $\mps$ with maximally twisted mass fermions
at four different values of the lattice spacing ranging from $0.051\,
\mathrm{fm}$ to $0.1\, \mathrm{fm}$ and for various values of the
quark mass corresponding to values for the pseudo scalar mass from
$280\,\mathrm{MeV}$ to $650\,\mathrm{MeV}$. In addition to $\fps$ and
$\mps$ we have determined the quark mass renormalisation constant
$Z_\mu=1/Z_\mathrm{P}$ (at the three largest values of the lattice
spacing only) and the hadronic length scale $r_0/a$.

The Sommer parameter $r_0$~\cite{Sommer:1993ce} is used as a scaling variable in our
investigations, but not to set the physical scale. We have employed
different ans{\"a}tze for extrapolating $r_0/a$ to the chiral
limit. While our data cannot discriminate between these different
ans{\"a}tze, it turns out that all of our final physical results are
independent of the exact way $r_0/a$ is chirally extrapolated. This is
due to the fact that the physical scale is eventually set by the pion
decay constant $f_\pi$. The values of $r_0^\chi/a$ we used for the
continuum-limit scaling analysis were obtained by chirally
extrapolating $r_0/a$ linearly in $(a\mu_q)^2$ at every value of the
lattice spacing separately, as explained in
section~\ref{sec:ens}. Later on, in the context of combined chiral
fits, we also allow $r_0/a$ to have an additional $a\mu_q$ dependence,
as discussed in section~\ref{sec:comb}.

\begin{figure}[t]
  \centering
  \subfigure[\label{fig:r0fps}]%
  {\includegraphics[width=0.45\linewidth]{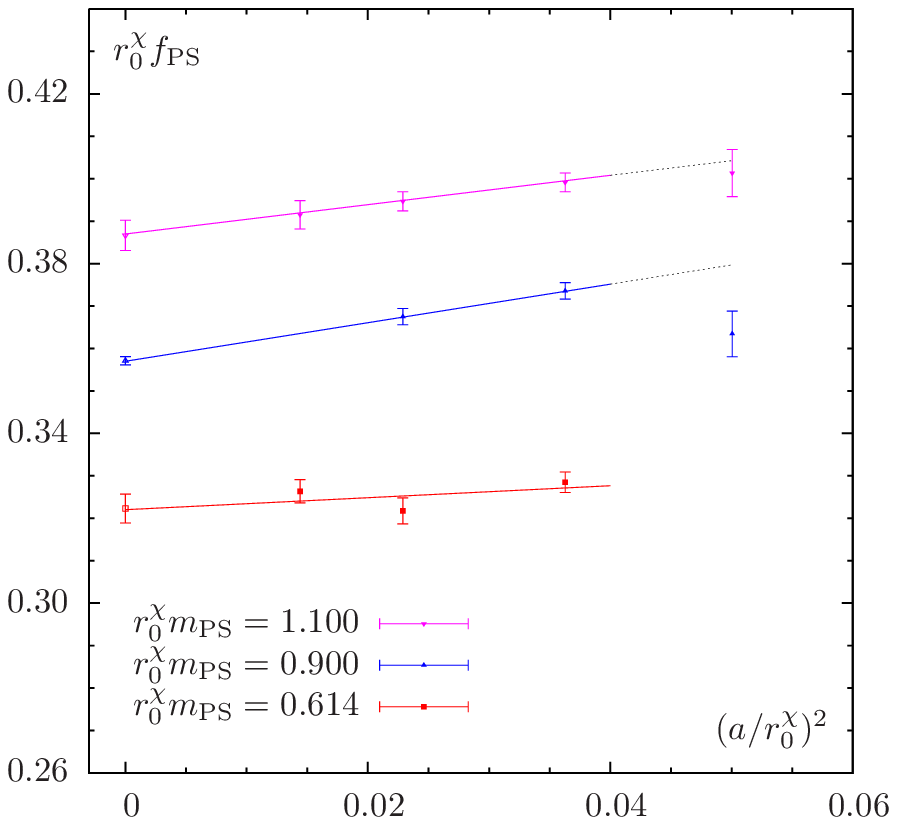}}
  \quad
  \subfigure[\label{fig:r0mps}]%
  {\includegraphics[width=0.445\linewidth]{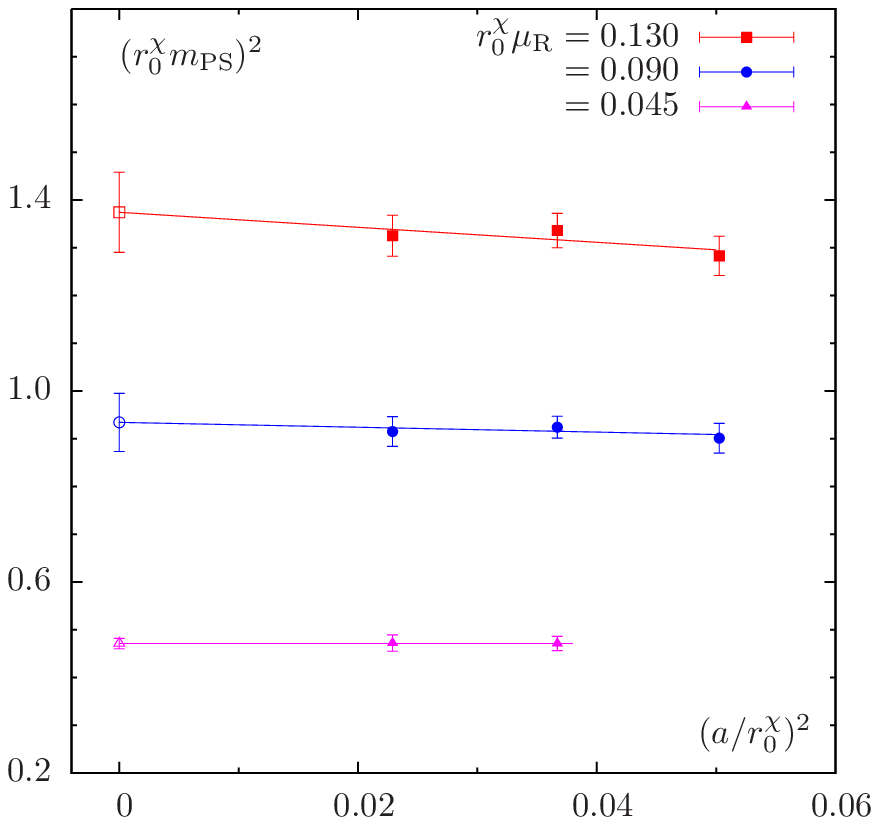}}  
  \caption{Scaling in finite, fixed volume for $r^\chi_0\fps$ at fixed values of
    $r^\chi_0\mps$ (a) and for $(r^\chi_0\mps)^2$ at fixed values of $r^\chi_0\mu_R$
    (b). In (b) we cannot include data at $\beta=4.2$ due to the
    missing value of the renormalisation factor $Z_\mathrm{P}$.}
  \label{fig:cont}
\end{figure}

Leading lattice artefacts in our data are expected to be of order
$a^2$, as discussed previously. This can be checked by extrapolating a
physical quantity at fixed physical situation to the continuum
limit. We show two such examples in figure~\ref{fig:cont}. In the left
panel we show $r^\chi_0\fps$ as a function of $(a/r^\chi_0)^2$ at
fixed value of $r^\chi_0\mps$. In order to match the values of
$r^\chi_0\mps$ at each value of $r^\chi_0/a$ and to fix the volume to
$r_0^\chi\cdot L=5$ we
had to perform short inter- or extra-polations. The straight lines are
linear fits in $\left( a/r^\chi_0\right)^2$ to the corresponding data,
with the data at the largest value of the lattice spacing not being
included in the fit. It is clearly visible that the lattice artefacts
appear to scale linearly in $a^2$ and that their overall size is
small.

In the right panel we show the scaling of $r^\chi_0\mps$ as a function of
$(a/r^\chi_0)^2$ at fixed values of the renormalised quark mass
$r^\chi_0\mu_R$, again at fixed, but finite volume. We conclude that 
also the charged pseudo scalar mass has only small lattice artefacts. 
This observation will become particularly important when we shall 
discuss later the mass difference between charged and neutral
pseudo scalar mesons. Note that we do not yet
have the value of $Z_\mathrm{P}$ for the smallest value of the lattice
spacing which is hence left out in fig.~\ref{fig:r0mps}.

The dependence of $m_\mathrm{PS}$ and $f_\mathrm{PS}$ on the
renormalised quark mass and volume can be described by chiral
perturbation theory
($\chi$PT)~\cite{Weinberg:1978kz,Gasser:1983yg,Gasser:1985gg}. The
residual lattice artefacts of 
order $a^2$ can also be included in the analysis.  The corresponding
formulae can be found in equation (\ref{eq:fmps}) below. We fit these
formulae to our data in order to extract the parameters of the $N_f=2$
chiral Lagrangian, i.e. the low energy constants and some derived
quantities. Moreover, we can use these fits to calibrate our lattices
by determining the value of the renormalised quark mass
$r^\chi_0\mu_R$ where the ratio $\mps/\fps$ assumes its physical value
(i.e. $m_\pi/f_\pi$) and set $\fps=f_\pi=130.7\, \mathrm{MeV}$
there,\,\footnote{Note that the recent update~\cite{Amsler:2008zzb}
  for $f_\pi=130.4(2)\, \mathrm{MeV}$ does not affect our results
  within the errors we quote.} as done in
ref.~\cite{Boucaud:2007uk}. Hence, $f_\pi$ is used in this paper to
set the scale.

\begin{table}[t!]
  \centering
  \begin{tabular*}{.85\linewidth}{@{\extracolsep{\fill}}lrrr}
    \hline\hline
    $\Bigl.\Bigr.$ & median & statistical & systematic \\
    \hline\hline
    $m_{u,d}\ [\mathrm{MeV}]$                            & $3.54$ & $(19)$ & $(+16-17)$ \\  
    $\bar\ell_3$                                         & $3.50$ & $(9)$ & $(+9-30)$ \\  
    $\bar\ell_4$                                         & $4.66$ & $(4)$ & $(+4-33)$ \\  
    $f_0\ [\mathrm{MeV}]$                                & $121.5$ & $(0.1)$ & $(+1.1-0.1)$ \\
    $B_0\ [\mathrm{MeV}]$                                & $2638$ & $(149)$ & $(\pm132)$ \\ 
    $r_0\ [\mathrm{fm}]$                                 & $0.420$ & $(9)$ & $(+10-11)$ \\ 
    $|\Sigma|^{1/3}\ [\mathrm{MeV}]$                       & $270$ & $(5)$ & $(+3-4)$ \\
    $f_\pi/f_0$                                          & $1.0755$ & $(6)$ & $(+8-94)$ \\
    $r^\chi_0/a(\beta=3.90)$                             & $5.32$ & $(5)$ & $(\pm0)$ \\ 
    $r^\chi_0/a(\beta=4.05)$                             & $6.66$ & $(6)$ & $(\pm0)$ \\ 
    $a(\beta=3.90)\ [\mathrm{fm}]$                       & $0.079$ & $(2)$ & $(\pm2)$ \\
    $a(\beta=4.05)\ [\mathrm{fm}]$                       & $0.063$ & $(1)$ & $(+1-2)$ \\
    $Z_\mathrm{P}(\beta=3.90)$                           & $0.434$ & $(8)$ & $(+4-2)$ \\
    $Z_\mathrm{P}(\beta=4.05)$                           & $0.452$ & $(9)$ & $(+3-9)$ \\ 
    \hline
   \end{tabular*}
  \caption{Summary of fit results, determined from the weighted
    distribution as explained in the text. The first error is of
    statistical origin while the second, the asymmetric one, accounts
    for the systematic uncertainties. $B_0$, $\Sigma$ and $m_{u,d}$
    are renormalised in the $\overline{\mathrm{MS}}$ scheme at the
    renormalisation scale $\mu = 2\, \mathrm{GeV}$, as the values of
    $Z_\mathrm{P}$ are in the $\overline{\mathrm{MS}}$ scheme at scale
    $2\, \mathrm{GeV}$. Note that for the results listed here only
    data at $\beta=3.9$ and $\beta=4.05$ have been used.  The scale is
    set by $f_\pi = 130.7\, \mathrm{MeV}$ as done in
    ref.~\cite{Boucaud:2007uk}. For a comparison to other recent lattice
    results we refer the reader to ref.~\protect{\cite{Scholz:2009yz}}.}
  \label{tab:results}
\end{table}

The results of these fits can be found in table \ref{tab:results}. We
give statistical and systematic errors separately, the systematic one
being asymmetrical. The results are obtained by performing
$\mathcal{O}(80)$ fits, which differ in fit-range, finite size
correction formulae and in the order of $\chi$PT. The final result is
obtained as the median of the corresponding weighted distribution over
all fits. The statistical error is determined using the bootstrap
method with 1000 samples. The systematic uncertainty is estimated from
the 68\% confidence interval of the weighted distribution. For details
we refer to the remaining sections of the paper, in particular to 
appendix~\ref{sec:fitdetails}. For a comparison to other lattice
results we refer the reader to ref.~\cite{Scholz:2009yz}.

The results quoted in table~\ref{tab:results} are obtained from the
two intermediate values of the lattice spacing only, since at the
smallest available lattice spacing $Z_\mathrm{P}$ has not yet been
computed and the largest value of $a$ is not considered, since there
our condition for tuning to maximal twist might not have been
fulfilled accurately enough, which may affect in particular the observables $\fps$ and
$\mps$ (at the smallest quark mass values) considered in this
paper. However, including data at the smallest (by fitting
$Z_\mu=Z_\mathrm{P}^{-1}$) and the largest value of the lattice
spacing gives results which are compatible within errors with the
numbers shown in table~\ref{tab:results}, as can be seen from a
direct comparison in table~\ref{tab:results2}.


\section{Ensemble details, $r_0$, $a^2$ and finite size effects}
\label{sec:ens}

Before discussing the chiral perturbation theory fits to
$m_\mathrm{PS}$ and $f_\mathrm{PS}$ in section~\ref{sec:comb} in
detail we shall address in this section a number of issues which are
needed as preparatory steps. These are the determination of the
hadronic scale $r_0/a$, the tuning to maximal twist, the size of the
$\mathrm{O}(a^2)$ corrections (including isospin breaking effects) and
finite size effects.

\subsection{Ensemble Details}

In table~\ref{tab:setup} we summarise the various $N_f=2$ ensembles
generated by the ETM collaboration. We remark here that all the
configurations produced by ETMC are available on the international
lattice data grid (ILDG) (see ref.~\cite{Yoshie:2008aw} and references
therein), and the configurations are made public after
publication of this paper. We have simulations at four
different values of the inverse gauge coupling
$\beta=3.8$, $\beta=3.9$, $\beta=4.05$ and $\beta=4.2$. 
The corresponding values of
the lattice spacing, obtained from chiral fits as detailed later, are  
$a\approx 0.1\, \mathrm{fm}$,
$a\approx 0.079\, \mathrm{fm}$, 
$a\approx 0.063\, \mathrm{fm}$ and 
$a\approx 0.051\, \mathrm{fm}$, 
respectively. For each value of $\beta$ we have several 
values of the bare twisted mass parameter $a\mu_q$,
chosen such that the ensembles cover a range of 
pseudo scalar masses between $280$ and $650\, \mathrm{MeV}$. 

The physical box length $L$ of most of the simulations at $\beta=3.9$,
$\beta=4.05$ are roughly equal and around $L\approx 2\, \mathrm{fm}$,
while the volume at $\beta=3.8$ is slightly larger. At $\beta=4.2$,
the lattice size ranges from $L\approx 1.7\, \mathrm{fm}$ to $2.5 \,
\mathrm{fm}$. For all the values of $\beta$, we demand the condition $\mps L
\gtrsim 3$. Note that we have carried out several simulations at
different physical volumes for otherwise fixed parameters in order to
study finite size effects (FSE).

The simulation algorithm used to generate these ensembles is a Hybrid
Monte Carlo algorithm with multiple time scales and mass
preconditioning. It is described in detail in
ref.~\cite{Urbach:2005ji} and one implementation described in
ref.~\cite{Jansen:2009xp} is freely available. 
Its performance for the current set-up is discussed in
ref.~\cite{Urbach:2007rt}. In table~\ref{tab:setup} we provide the
values of the actual trajectory 
length $\tau$ used in the simulations. For each value of $\beta$ and
$a\mu_q$ we have produced  around $5000$ equilibrated trajectories in
units of $\tau=0.5$. In all cases we allowed for at least $1500$
trajectories for equilibration (again in units of $\tau=0.5$). Some of
the ensembles have been produced with a trajectory length $\tau=1$
following the suggestions of ref.~\cite{Meyer:2006ty}.

\begin{table}[t!]
  \centering
  \begin{tabular*}{1.0\textwidth}{@{\extracolsep{\fill}}lccccccc}
    \hline\hline
    Ensemble & $(L/a)^3\times T/a$ & $\beta$ & $a\mu_q$ & $\kappa_\mathrm{crit}$ &
    $\tau_\mathrm{int}(P)$ & $\tauint(am_\mathrm{PS})$ & $\tau$\\
    \hline\hline
    $A_1$ & $24^3\times 48$ & $3.8$ & $0.0060$ & $0.164111$ & $190(44)$
    & $8(2)$ & $1.0$\\
    $A_2$ &  & & $0.0080$ & & $172(80)$ & $10(2)$ & $1.0$\\
    $A_3$ &  & & $0.0110$ & & $130(50)$ & $6(1)$ & $1.0$\\
    $A_4$ &  & & $0.0165$ & & $40(12)$ & $6(1)$ & $1.0$\\
    $A_5$ & $20^3\times 48$ & & $0.0060$ &  & $250(100)$
    & $5(1)$ & $1.0$\\
    \hline
    $B_1$ & $24^3\times 48$ & $3.9$ & $0.0040$ & $0.160856$ & $47(15)$
    & $7(1)$ & $0.5$\\
    $B_2$ &  & & $0.0064$ &  & $23(7)$
    & $17(4)$ & $0.5$ \\
    $B_3$ &  & & $0.0085$ &  & $13(3)$ & $10(2)$ & $0.5$\\
    $B_4$ &  & & $0.0100$ &  & $15(4)$ & $7(2)$ & $0.5$\\
    $B_5$ &  & & $0.0150$ &  & $30(8)$ & $20(6)$ & $0.5$\\
    $B_6$ & $32^3\times 64$ &  & $0.0040$ & & $37(11)$
    & $2.8(3)$ & $0.5$ \\
    $B_7$ &  &  & $0.0030$ & & $51(19)$
    & $7(1)$ & $1.0$ \\
    \hline
    $C_1$ & $32^3\times 64$ & $4.05$ & $0.0030$ & $0.157010$ & $18(4)$ &
    $7(1)$ & $0.5$\\
    $C_2$ & & & $0.0060$ & & $10(2)$ & $9(2)$ & $0.5$ \\
    $C_3$ & & & $0.0080$ & & $13(3)$ & $7(1)$ & $0.5$\\
    $C_4$ & & & $0.0120$ & & $5(1)$ & $4.8(6)$ & $0.5$\\
    $C_5$ & $24^3\times 48$ &  & $0.0060$ & & $12(2)$ &
    $11(1)$ & $1.0$\\
    $C_6$ & $20^3\times 48$ &  & $0.0060$ & & $10(2)$ &
    $7(1)$ & $1.0$\\
    \hline
    $D_1$ & $48^3\times 96$ & $4.2$ & $0.0020$ & $0.154073$ & $13(2)$ & $\leq8$ & $1.0$ \\
    $D_2$ & $32^3\times 64$ &       & $0.0065$ &            & $6(1)$  & $\leq8$ & $1.0$ \\
    \hline
  \end{tabular*}
  \caption{Summary of ensembles generated by ETMC. We
    give the lattice volume $L^3\times T$  
    and the values of the
    inverse coupling $\beta$, the twisted mass parameter $a\mu_q$, the
    critical hopping parameter $\kappa_\mathrm{crit}$ as determined at $\mu_{q,\mathrm{min}}$
    (i.e. the value of $a\mu_q$ appearing in the same row as $\kappa_\mathrm{crit}$)
    and the
    trajectory length $\tau$. 
    The values of the lattice spacing that correspond to the four values 
    of $\beta$ are $a\approx 0.1\, \mathrm{fm}$ ($\beta=3.8$), $a\approx0.079\, \mathrm{fm}$ ($\beta=3.9$), 
    $a\approx0.063\, \mathrm{fm}$ ($\beta=4.05$) and $a\approx0.051\, \mathrm{fm}$ ($\beta=4.2$).
    In addition
    we provide values for the integrated autocorrelation time of two
    typical quantities, the plaquette $P$ and the pseudo scalar mass
    $am_\mathrm{PS}$, in units of $\tau=0.5$. We refer to ref.~\cite{Boucaud:2008xu} 
    for details on the determination of the autocorrelation time.}
  \label{tab:setup}
\end{table}

\subsection{Hadronic scale parameter $r_0$}

In order to be able to compare results at different values of the
lattice spacing it is convenient to use the hadronic scale $r_0$
\cite{Sommer:1993ce}. It is defined via the force between static
quarks and can be measured to high accuracy in lattice QCD
simulations. For details on our procedure to determine $r_0/a$ we 
refer to ref.~\cite{Boucaud:2008xu}.
Note that in this work we use $f_\pi$ to set the scale and employ
$r_0$ only for a continuum limit scaling analysis. In other words
$r_0^{\chi}$ is used only as an intermediate reference quantity, which
is finally eliminated in favour of $f_\pi$.

\begin{figure}[t]
  \centering
    \subfigure[\label{fig:r039}]%
  {\includegraphics[width=0.46\linewidth]{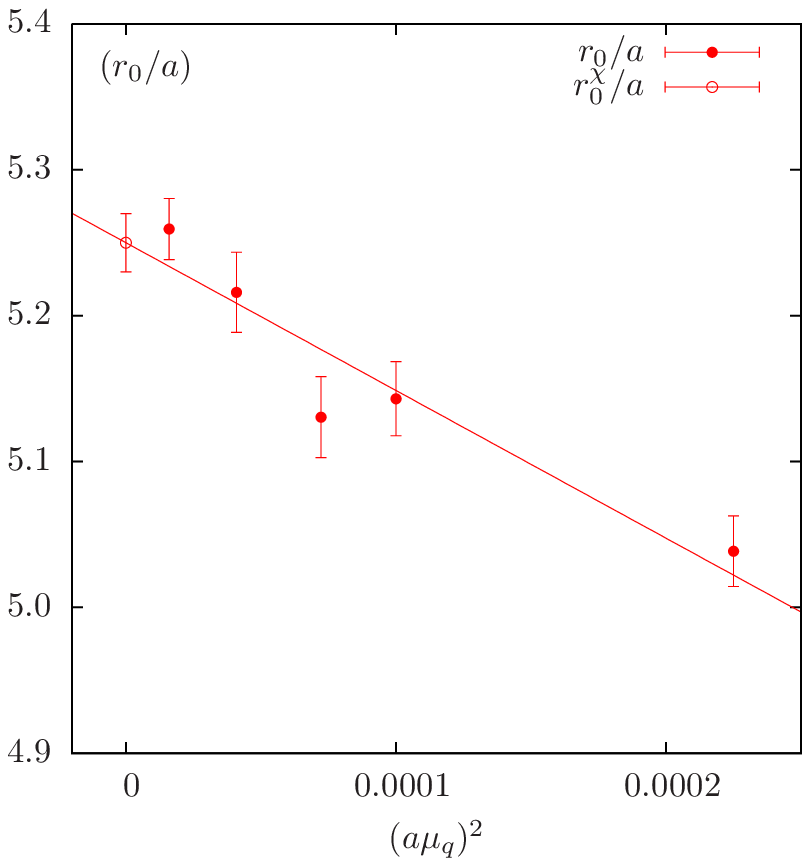}}
  \quad
  \subfigure[\label{fig:r0405}]%
  {\includegraphics[width=0.44\linewidth]{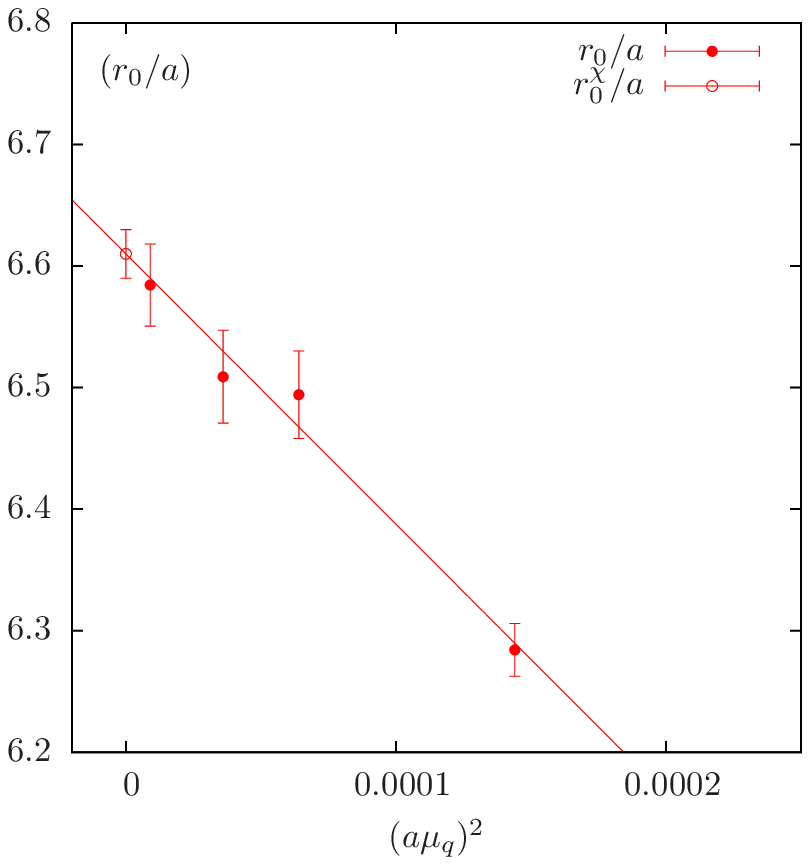}}
  \caption{$r_0/a$ as a function of $(a\mu_q)^2$ for
    (a) $\beta=3.9$ and (b) $\beta=4.05$. The lines represent a
    linear extrapolation in $(a\mu_q)^2$ to the chiral limit. Note
    that we have always used the largest available volume for a given
    value of $a\mu_q$, see table~\ref{tab:setup}. 
  }
  \label{fig:r0}
\end{figure}

In fig.~\ref{fig:r0} we show $r_0/a$ as a function of $(a\mu_q)^2$ for
the examples of $\beta=3.9$ and $\beta=4.05$. The mass dependence
appears to be rather weak and a linear fit in $\mu_q^2$ describes
the data well as has been also discussed in
refs.~\cite{Boucaud:2007uk,Boucaud:2008xu}. However, also a linear
dependence on $a\mu_q$ cannot be excluded -- due to
possible effects of spontaneously broken chiral symmetry in
$r_0$. This dependence describes in the case of $\beta=3.90$ the data
even better than the quadratic ansatz. We shall therefore
conservatively include both terms in our final analysis, i.e.\ the
chiral fits discussed in section~\ref{sec:comb}. 

For tuning to maximal twist as well as for the continuum-limit scaling
analyses presented in this section -- in 
particular in figures~\ref{fig:cont}, \ref{fig:mpcac} and \ref{fig:tm}
-- we shall use the value of $r_0^\chi/a$ from a linear extrapolation
in $(a\mu_q)^2$ to the chiral limit. A linear
extrapolation in $(a\mu_q)^2$ yields $r_0^\chi/a=4.47(6)$ at
$\beta=3.8$, $r_0^\chi/a=5.25(2)$ at $\beta=3.9$, $r_0^\chi/a=6.61(2)$
at $\beta=4.05$ and $r_0^\chi/a=8.33(5)$ at $\beta=4.2$. The
statistical accuracy for $r^\chi_0/a$ is about $0.5\%$. It is
important to notice that the ratio
$[r_0^\chi/a(\beta=3.9)]/[r_0^\chi/a(\beta=4.05)]$ is within the
errors independent of the extrapolation procedure. This leads us to
expect rather little influence of the $r_0/a$ extrapolation strategy
on physical quantities.

\subsection{Tuning to Maximal Twist}

Let us briefly discuss our strategy to tune to maximal twist.                   
It corresponds to tune $am_0$ to a critical value $am_\mathrm{crit}$ 
such that the PCAC quark mass defined in eq.~(\ref{eq:mpcac}) 
vanishes. In particular, we determine the value of $am_\mathrm{crit}$
at the lowest available value of $a\mu_{q,\mathrm{min}} \ll
a\Lambda_\mathrm{QCD}$ at each $\beta$-value. Furthermore, we demand
that the ratio $| Z_\mathrm{A} am_\mathrm{PCAC}/a\mu_{q,\mathrm{min}}| \lesssim 0.1$ and
also that its error $\Delta(Z_\mathrm{A} am_\mathrm{PCAC}/a\mu_{q,\mathrm{min}}) \lesssim  0.1$.
These criteria and their justification for tuning to maximal
twist have been discussed in ref.~\cite{Boucaud:2008xu}.
An additional condition is that the value of $a\mu_{q,\mathrm{min}}$
corresponds to a fixed physical situation at all values of the lattice
spacing. We have chosen to use $m_\mathrm{PS}
\simeq 300\mathrm{MeV}$ to define the value of 
$a\mu_{q,\mathrm{min}}$. Considering $\beta=3.9$,
$\beta=4.05$ and $\beta=4.2$, it was possible to perform this tuning task 
with two or three tuning runs for each lattice spacing.
Obeying these conditions leads to very 
small $\mathrm{O}(a^2)$ effects in the physical observables considered
in this paper, with the only exception 
of the neutral pseudo scalar mass, which will be discussed
below.  

\begin{figure}[t]
  \centering
  \includegraphics[width=0.8\linewidth]{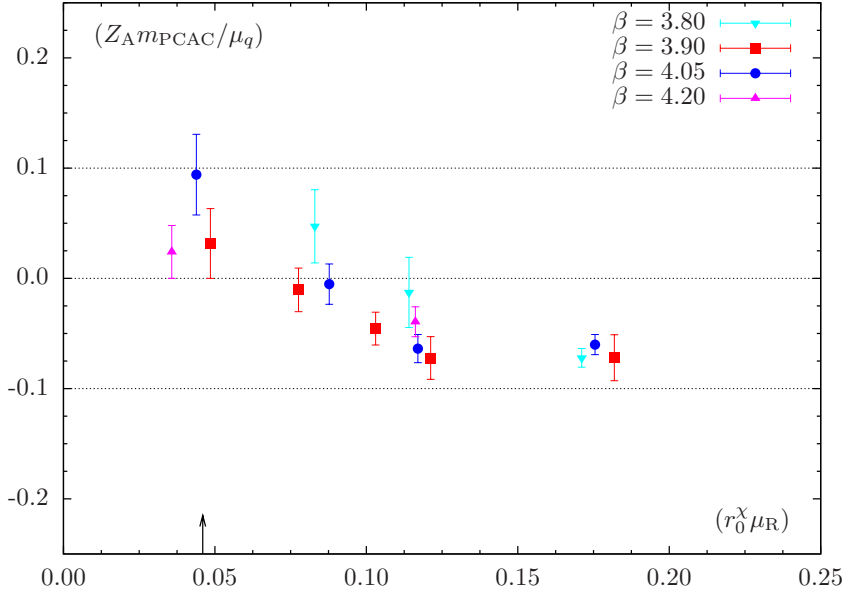}
  \caption{Renormalised ratio of the PCAC quark mass over the twisted
    mass against the renormalised twisted mass
    $\mu_\mathrm{R}=\mu_q/Z_\mathrm{P}$ at the four values of
    $\beta$. The statistical uncertainties on $Z_\mathrm{P}$ and
    $Z_\mathrm{A}$ are not included.  The data at $\beta=4.2$ have
    been included by estimating the renormalisation constants as
    described in the text. At $\beta=3.8$, the data for the lightest
    quark mass has not been included for the reasons explained in the
    text. The band indicates our condition for tuning to maximal
    twist, which is clearly achieved to a good precision. The arrow
    indicates the value of $r_0^\chi\mu_\mathrm{R}$ where we tuned the
    PCAC mass to zero.}
  \label{fig:mpcac}
\end{figure}

A special case are the simulations at 
$\beta=3.8$. As can be seen in table~\ref{tab:setup} at the smallest value
of $a\mu_q=0.006$ we encounter very large autocorrelation times for 
the plaquette. Since the autocorrelation time for the PCAC quark mass
is of the same order of magnitude as the one of the plaquette, at this value of $a\mu_q$ it is 
questionable whether $am_\mathrm{PCAC}$ was reliably determined since 
the effective statistics is small. Hence, it is unclear whether a tuning
to maximal twist fulfilling our conditions has been achieved at
$\beta=3.8$. A clarification of this point would have required
substantially larger  statistics. As a consequence, we will leave out
the  data at $\beta=3.8$ in our main analysis. The $\beta=3.8$ data
will only be discussed in the context of systematic uncertainties.

The measurements of $\mpcac$ from all the ensembles are collected in
Appendix~\ref{sec:data}.  In figure~\ref{fig:mpcac} we show the
properly renormalised ratio of the PCAC quark mass $\mpcac$ over the
twisted mass $\mu_q$ (the renormalisation factor of $Z_\mathrm{P}$
cancels in this ratio) against the renormalised twisted mass,
$\mu_\mathrm{R}=\mu_q/Z_\mathrm{P}$, in units of $r_0^\chi$, for the
four values of $\beta$.  The renormalisation factors $Z_\mathrm{P}$,
taken here at a scale of $2$\,GeV, were determined using the RI'-MOM
scheme \cite{Martinelli:1994ty,Dimopoulos:2007fn}. Note that for
$\beta=4.2$ the renormalisation factors are not available
yet. However, we have estimated their values by fitting $Z_\mathrm{P}$
and by estimating the value of $Z_A=0.75$ from the $\beta$-dependence
of its known values at smaller $\beta$. Since these renormalisation
constants enter as $\mathcal{O}(1)$ multiplicative pre-factors, a small change
in their values will not significantly change the data in
figure~\ref{fig:mpcac}. Therefore our main conclusion about the
quality of the tuning will be un-altered if the actual numbers for the
renormalisation constants should come out slightly differently in the
final analysis.

As can be seen from fig.~\ref{fig:mpcac} 
our condition for tuning to maximal twist is satisfied with a good
statistical precision.  The fact that at the three largest values of
$\beta$ the renormalised quark mass computed at the tuning twisted
mass parameter $a\mu_{q,\mathrm{min}}$ basically agree (this is
illustrated by the arrow in fig.~\ref{fig:mpcac}) shows in addition
that the physical situation has indeed been fixed.  At $\beta=3.8$,
due to the previously mentioned problem of the long autocorrelation
time at the lowest quark mass, we include in fig.~\ref{fig:mpcac} only
the ensembles for which a reliable estimate of the error on $\mpcac$
was possible. At all $\beta$-values for values of
$a\mu_q>a\mu_{q,\mathrm{min}}$ we observe in $a\mpcac$ (small)
deviations from zero. This $\mathcal{O}(a)$ cut-off effect will modify
\emph{only} the $\mathcal{O}(a^2)$ lattice artefacts of physical
observables.

\subsection{$\mathcal{O}(a^2)$ effects}

In this section, we shall discuss the lattice spacing effects on the
decay constant $f_\mathrm{PS}$ and the mass $m_\mathrm{PS}$ of the
(charged) pseudo scalar meson. Starting with $f_\mathrm{PS}$, we will
use the data at $\beta=3.9$, $\beta=4.05$ and $\beta=4.2$. In order to
compare the data at these values of $\beta$, we have fixed three
reference values of the pseudo scalar mass to
$r^\chi_0m_\mathrm{PS}=0.614$, $r^\chi_0\mps=0.90$ and
$r^\chi_0m_\mathrm{PS}=1.10$. The corresponding values for
$af_\mathrm{PS}$ were then obtained by small interpolations, which we
have chosen to be linear in $\mps^2$. In addition, we also corrected for the very
small differences in the physical volume for the $\beta$ values used
such that all data were scaled to the same physical volume
$r_0^\chi\cdot L=5.0$. This has been achieved by applying the relevant
formulae from $\chi$PT as detailed below.

As has already been discussed in section~\ref{sec:mainres}, in 
fig.~\ref{fig:r0fps} (cf. page~\pageref{fig:r0fps}) we show the pseudo
scalar decay constant, brought  
to common reference points $r^\chi_0m_\mathrm{PS}$ and volumes,  
as a function of $(a/r^\chi_0)^2$.
In our earlier work, where we had only two values of the lattice spacing
\cite{Urbach:2007rt,Dimopoulos:2007qy,Dimopoulos:2008sy}, we detected only 
very small lattice spacing artefacts when comparing $\beta=3.9$ and $\beta=4.05$. 
This observation is confirmed: in fig.~\ref{fig:r0fps} we now have added, 
for two values of $a\mu_q$, a smaller lattice spacing of $a=0.051\
\mathrm{fm}$ and we can clearly observe that the scaling is as
expected linear in $a^2$ with only a small slope such that no 
large cutoff effects are visible. Note that also for the nucleon 
mass the scaling violations are very small (see
refs.~\cite{Alexandrou:2008tn,Alexandrou:2009qu}).

We also show $r^\chi_0f_\mathrm{PS}$ for $\beta=3.8$ in
fig.~\ref{fig:r0fps} at least for those points where we are able to
interpolate to the reference values of
$r^\chi_0m_\mathrm{PS}$. Although the linear continuum extrapolation
is performed from the results at $\beta=3.9$, $\beta=4.05$ and
$\beta=4.2$, as can be seen, the data at $\beta=3.8$ are consistent
with this linear behaviour. This indicates that the lattice artefacts
are indeed small.

In fig.~\ref{fig:r0mps} we show similarly the continuum
extrapolation of $(r^\chi_0\mps)^2$ for three fixed values of the
renormalised quark mass $r^\chi_0\mu_\mathrm{R}=0.045, 0.09, 0.13$. In this
figure we cannot include data at the smallest lattice spacing, due to
the missing $Z_\mathrm{P}$-factor. Instead we include data from $\beta=3.80$ in
the continuum extrapolation to show that 
the scaling looks very flat. In conclusion also for $(r^\chi_0\mps)^2$
residual scaling violations are small.

\subsubsection{Isospin Breaking Effects}

A peculiar $\mathrm{O}(a^2)$ effect in 
the twisted mass formulation of lattice QCD is the breaking 
of isospin symmetry at any non-zero value of the lattice spacing. 
Investigations of these effects deserve therefore special attention. 
In the quenched approximation the difference 
$(r_0^\chi)^2((m_\mathrm{PS}^\pm)^2 -(m_\mathrm{PS}^0)^2)$ between 
the neutral $m_\mathrm{PS}^0$ and the charged 
$m_\mathrm{PS}^\pm$ pseudo scalar masses, although fully 
compatible with 
the expected $\mathrm{O}(a^2)$ behaviour, turned out to be 
significant \cite{Jansen:2005cg}. 

The determination of 
$m_\mathrm{PS}^0$ requires the computation of disconnected diagrams 
rendering its determination difficult. Nevertheless, following 
the techniques described in ref.~\cite{Boucaud:2008xu}, we were able 
to compute $\mps^0$ (and other neutral quantities) with a 
reasonable precision. The results for $m_\mathrm{PS}^0$ and 
$m_\mathrm{V}^0$ can be found in
table~\ref{tab:mpi0} in the appendix.

We show in fig.~\ref{fig:tm} the mass splitting between the charged 
and neutral pseudo scalar mesons, i.e. 
\begin{equation}
  \label{eq:split}
  (r_0^\chi)^2\ \left[(\mps^\pm)^2 -(\mps^0)^2\right]
\end{equation}
as a function of $(a/r_0^\chi)^2$.  As can be seen, also for $N_f=2$
dynamical quark flavours the isospin breaking effects are large,
although smaller than observed in the quenched approximation and of
opposite sign, cf. ref.~\cite{Jansen:2005cg}, meaning that in the case
of dynamical quarks considered here the neutral pseudo scalar meson
turns out to be lighter than the charged one. This observation is
consistent with the observed first order phase transition
\cite{Farchioni:2004us,Farchioni:2005tu} and with the corresponding
scenario of ref.~\cite{Sharpe:2004ps}.

One very important observation is that the large cutoff effect in the mass difference
between the neutral and charged pseudo scalar masses is dominated by 
discretisation effects in the 
neutral pseudo scalar mass. This is demonstrated in fig.~\ref{fig:r0mps} 
which shows the charged pseudo scalar mass as a function of $(a/r_0^\chi)^2$
for fixed values of $r_0^\chi\mu_R$. As can be seen, the lattice spacing 
effects are very small, as discussed above.

\begin{table}[t]
  \centering
  \begin{tabular*}{0.7\linewidth}{@{\extracolsep{\fill}}lrrr}
    \hline\hline
    & $\beta$ & $a\mu_q$ & $R_O$ \\
    \hline\hline
    $f_\mathrm{PS}$ & $3.90$ & $0.004$ & $0.04(06)$ \\
    & $4.05$ & $0.003$ & $-0.03(06)$ \\ 
    $m_\mathrm{V}$  & $3.90$ & $0.004$ & $0.02(07)$ \\
    & $4.05$ & $0.003$ & $-0.05(09)$ \\
    $m_\mathrm{V}f_\mathrm{V}$  & $3.90$ & $0.004$ & $-0.07(18)$ \\
    & $4.05$ & $0.003$ & $-0.31(29)$ \\
    $m_\Delta$      & $3.90$ & $0.004$ & $0.022(29)$  \\
    & $4.05$ & $0.003$ & $-0.004(45)$ \\
    \hline
  \end{tabular*}
  \caption{Comparison of some selected quantities for which an 
    isospin splitting can occur for twisted mass fermions. $R_O$ denotes the 
    relative size of the splitting, see also ref.~\cite{Frezzotti:2007qv,Dimopoulos:2009qv}.}
  \label{tab:splitting}
\end{table}

Another observation is that, in the unitary sector, so far all other
investigated differences between corresponding quantities affected by
isospin violations are compatible with zero. In
table~\ref{tab:splitting} we have compiled the relative difference
$R_O=(O-O')/O$ for some selected observables and simulation points.
Here $O$ ($O'$) denotes the charged (neutral) quantity in the case of
mesons and $\Delta^+$ ($\Delta^{++}$) in the case of baryons. All of
them are well compatible with zero. However, some quantities, like the
vector meson decay constant $f_\mathrm{V}$, are rather noisy and hence
the result for this quantity is not really conclusive.

\begin{figure}[t]
  \centering
  {\includegraphics[width=0.45\linewidth]{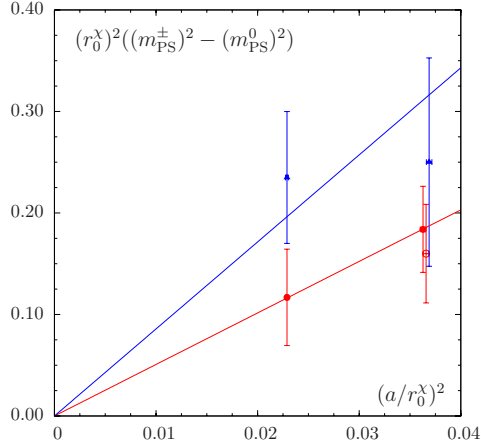}}\quad
  \caption{The difference of the squared charged and neutral pseudo scalar
    masses as a function of $a^2$ in the $N_f=2$ twisted mass
    formulation of lattice QCD at two different values of the charged
    pseudo scalar mass. A significant $\mathrm{O}(a^2)$ lattice
    artefact is observed. The circles (triangles) correspond to a
    value of the charged pion mass of about $330$~($430$)\,MeV. The
    open circle is a larger physical volume. The lines are only to
    guide the eye and some points are slightly horizontally displaced
    for better visibility. Note that $r_0^\chi\mps^\pm$ was not held
    fixed for this plot, however, due to the large uncertainties on
    $\mps^0$ the picture should not significantly depend on this
    approximation.}
  \label{fig:tm}
\end{figure}

The ETM collaboration has been investigating the question 
of the large cutoff effects in the neutral pseudo-scalar mass 
\cite{Frezzotti:2007qv,Dimopoulos:2009qv,Munster:2004am} also theoretically. An
analysis \`a la Symanzik of the charged and the neutral pseudo scalar
meson masses leads to the formulae  
\begin{equation}
  \label{eq:sym}
  \begin{split}
    (m_\mathrm{PS}^0)^2 &= m_\pi^2 + a^2\zeta_\pi +
    \mathcal{O}(a^2m_\pi^2,a^4)\\ 
    (m_\mathrm{PS}^\pm)^2 &= m_\pi^2 +
    \mathcal{O}(a^2m_\pi^2,a^4)\\ 
  \end{split}    
\end{equation}
which show that the difference
$(m_\mathrm{PS}^0)^2-(m_\mathrm{PS}^\pm)^2$  
is given by the term proportional to $\zeta_\pi$ (for a discussion in
the framework of twisted mass Wilson $\chi$PT see
ref.~\cite{Scorzato:2004da,Sharpe:2004ny}). Here 
$\zeta_\pi\equiv\langle\pi^0|\mathcal{L}_6|\pi^0\rangle$ and
$\mathcal{L}_6$ is the dimension six term in the Symanzik 
effective Lagrangian.

The main result of the analysis of ref.~\cite{Frezzotti:2007qv,Dimopoulos:2009qv,Munster:2004am}
is that $\zeta_\pi$ is a large number which in the 
vacuum saturation approximation can be estimated 
to be proportional to $|\hat{G}_\pi|^2$, where 
$\hat{G}_\pi =\langle 0 |\hat{P}^3|\pi^0\rangle$. The latter matrix
element is numerically large: one finds
$|\hat{G}_\pi|^2/\Lambda_\mathrm{QCD}^4$ around
$20-25$~\cite{Frezzotti:2007qv,Dimopoulos:2009qv,Munster:2004am}. This
result provides a physical explanation for the large
$\mathcal{O}(a^2)$ effect in the neutral pseudo scalar
mass. Moreover, since it can be shown that $\zeta_\pi$ appears only in
the  neutral pseudo scalar mass (and kinematically related quantities), one also
finds a possible explanation of why all other splittings determined so
far turn out to be small.
We remark that in principle also a double insertion of the operator 
$\mathcal{L}_5$ can contribute to the $\mathcal{O}(a^2)$ lattice artefacts.
However, it turns out that these contributions are much smaller than 
those of $\mathcal{L}_6$ discussed here \cite{Dimopoulos:2009qv}.


\subsection{Finite Size Effects in $\fps$ and $\mps$}
\label{sec:fs}

In the following, we shall focus on the treatment of finite size
effects in $\fps$ and $\mps$ with the aim of assessing
the applicability of an effective description to be used, in
particular, to perform the very small volume corrections required to
bring the simulation points to a common finite volume, as needed for
the continuum-limit scaling analysis. Later on, in the context of the
chiral fits described in section~\ref{sec:comb}, we will apply the
finite size corrections formulae to bring our continuum data to
infinite volume and to check for systematic effects.

Our lattice data have been obtained with values of
$m_\mathrm{PS}L\gtrsim 3$. It is generally believed that in such a
situation the finite size effects are small and appear only at the per
cent level. However, our simulation data are on a level of precision
that such effects are certainly detectable and hence finite size
effects  can in general not be neglected. One possible way to control 
finite size effects is to use chiral perturbation theory ($\chi$PT) 
which provides an
effective description for the dependence of physical quantities on the
box size~\cite{Gasser:1986vb}. A completely independent analytical
description of finite size effects (FSE) of particle masses is
provided by the so called L{\"u}scher formula developed in
ref.~\cite{Luscher:1985dn}.

Being interested in $\mps$ and $\fps$ -- where the
applicability of $\chi$PT should be best justified -- we then have two
effective descriptions of FSE at hand: (i) $\chi$PT to next-to-leading
order (NLO) by Gasser and Leutwyler \cite{Gasser:1986vb}, which we shall 
denote with GL for short, and (ii) the
extended asymptotic L{\"u}scher formula of Colangelo, D\"urr and
H\"afeli~\cite{Colangelo:2003hf,Colangelo:2005gd}, which conveniently
combines the L{\"u}scher formula approach with $\chi$PT
and which we will refer to
as CDH in the following. Note that in pure $\chi$PT the finite size corrections
to $\mps$ are also known to 2-loop order~\cite{Colangelo:2006mp}.
However, it turns out that the difference between the CDH formula and
2-loop $\chi$PT is negligible~\cite{Colangelo:2006mp}.

Below, we compare the numerical data obtained by ETMC to 
the two different analytical descriptions of the FSE in $\mps$ and
$\fps$, which is possible since we have precise
values for these quantities for a number of ensembles such as  
$C_2$, $C_5$ and $C_6$ or $B_1$ and $B_6$ where all parameters 
but the lattice size are fixed.  
We remark that for all these ensembles $\mps L\geq 3$ holds, such that
the formulae from
refs.~\cite{Gasser:1986vb,Colangelo:2003hf,Colangelo:2005gd,Colangelo:2006mp}
should be applicable. This was also checked and discussed in
ref.~\cite{Urbach:2007rt}. 

Starting with the GL
ansatz, the finite size corrections  for $m_\mathrm{PS}$ and
$f_\mathrm{PS}$ read 
\begin{equation}
  \label{eq:fs}
  \begin{split}
    m_\mathrm{PS}(L) &=
    m_\mathrm{PS}\Bigl[1+\frac{1}{2}\xi\tilde{g}_1(\lambda)\Bigr]\equiv
    m_\mathrm{PS}\ K_m^{\mathrm{GL}}(L, \mps, f_0)\, ,\\
    f_\mathrm{PS}(L) &= 
    f_\mathrm{PS}\ \Bigl[1-2\xi\tilde{g}_1(\lambda)\Bigr]\ \equiv
    f_\mathrm{PS}\ K_f^\mathrm{GL}(L, \mps, f_0)\, ,
  \end{split}
\end{equation}
where
\begin{equation}
  \label{eq:xi}
  \xi = m_\mathrm{PS}^2/(4\pi f_0)^2\, ,\qquad\lambda = m_\mathrm{PS} L\, ,
\end{equation}
$\tilde{g}_1$ is a known function \cite{Gasser:1986vb} and the finite
size corrections $K_{m,f}^\mathrm{GL}$ depend apart from $L$ and
$m_\mathrm{PS}$ only on the unknown leading order low energy constant
$f_0$ representing the pseudo scalar decay constant in the chiral limit
(note that our normalisation is such that the physical value of the 
pseudo scalar decay constant is $f_\pi=130.7\, \mathrm{MeV}$). 

The CDH formulae have a very similar form
\begin{equation}
  \label{eq:cdh}
  \begin{split}
    \mps(L) &= \mps\Bigl[1- \sum_{n=1}^\infty
    \frac{m(n)}{2\sqrt{n}\lambda}\xi\{I_m^{(2)}+\xi
    I_m^{(4)}+\xi^2I_m^{(6)} + \mathcal{O}(\xi^3)\}\Bigr]\equiv m_\mathrm{PS}\ K_m^{\mathrm{CDH}}\, ,\\
    \fps(L) &= \fps\Bigl[1+ \sum_{n=1}^\infty
    \frac{m(n)}{\sqrt{n}\lambda}\xi\{I_f^{(2)}+\xi
    I_f^{(4)}+\xi^2I_f^{(6)} + \mathcal{O}(\xi^3)\}\Bigr]\equiv f_\mathrm{PS}\ K_f^\mathrm{CDH}\, ,\\
  \end{split}
\end{equation}
where the $I_{m,f}^{(i)}$ can be written in terms of basic integrals
as discussed in ref.~\cite{Colangelo:2005gd}, where also $m(n)$ is
tabulated. The authors of ref.~\cite{Colangelo:2005gd} provide
simplified formulae for the 
$I_{m,f}^{(i)}$, which we shall use in the following. Note that
$I_m^{(6)}$ is known while $I_f^{(6)}$ is not. However, $I_m^{(6)}$ is
numerically so small that we shall drop it, i.e. set it to zero, in
most of what follows. $I_f^{(6)}=0$ is always used.

The drawback of the CDH formulae compared to GL is that additional low
energy parameters are needed as an input\,\footnote{If $I_m^{(6)}$ is to be included
  there are also the parameters $\tilde r_{1-4}$ needed. 
For estimates of their values see ref.~\cite{Colangelo:2001df}}, namely: $\Lambda_1$, $\Lambda_2$,
$\Lambda_3$ and $\Lambda_4$. However,
there are estimates for all those parameters given in
ref.~\cite{Colangelo:2001df} and compiled in the form $\bar{\ell}_i =
\log(\Lambda_i^2/m_\pi^2)$ in table~\ref{tab:l12} for
convenience, which we shall rely on during the analysis presented in this section. In order to
convert the estimates in table~\ref{tab:l12} into lattice units we 
conventionally used in this analysis $r^\chi_0=0.45\, \mathrm{fm}$. Note that the choice
of scale here affects only the sub-leading contributions to the finite
size corrections and has a negligible effect, as was checked also by
direct inspection. We emphasise again that we are at this stage only
interested in an effective description of FSE with the goal to
understand whether the correction formulae are applicable in the
regime of our simulation parameters. 

\begin{table}[t!]
  \centering
  \begin{tabular}[t]{lc}
    \hline\hline
    $\bigl.\Bigr.i$ & $\bar{\ell}_i$ \\
    \hline\hline
    $1$ & $-0.4\pm0.6$ \\
    $2$ & $+4.3\pm0.1$ \\
    $3$ & $+2.9\pm2.4$ \\
    $4$ & $+4.4\pm0.2$ \\
    \hline
  \end{tabular}
  \caption{Values for $\bar{\ell}_{i}=\log(\Lambda_i^2/m_\pi^2)$ from
    table 2 of ref.~\cite{Colangelo:2005gd} used here to analyse FSE.}
  \label{tab:l12}
\end{table}

While the expansion in ref.~\cite{Colangelo:2001df} was performed in
terms of the squared pion mass, in ref.~\cite{Frezzotti:2008dr} the
same formulae have been obtained expanding in the quark mass
directly. We shall refer to the latter as CDH$^m$ and include it in
the following tests.

\begin{table}[t!]
  \centering
  \begin{tabular*}{1.\textwidth}{@{\extracolsep{\fill}}lccccccc}
    \hline\hline
    $\Bigl.\Bigr.$ & $L/a$ & $a\mps(L)$ & $a\mps^\mathrm{GL}(\infty)$ & 
    $a\mps^\mathrm{CDH,6}(\infty)$ & $a\mps^\mathrm{CDH}(\infty)$ &
    $a\mps^\mathrm{CDH^{m}}(\infty)$ \\
    \hline\hline
    $B_1$ & $24$ & $0.1362(7)$ & $0.1354(7)$ & $0.1348(7)$ & 
    $0.1350(7)$ & $0.1346(7)$ \\
    $B_6$ & $32$ & $0.1338(2)$ & $0.1336(2)$ & $0.1335(2)$ & 
    $0.1336(2)$ & $0.1335(2)$ \\
    \hline
    $C_6$ & $20$ & $0.1520(15)$ & $0.1492(15)$ & $0.1448(15)$ &
    $0.1464(15)$ & $0.1430(15)$ \\
    $C_5$ & $24$ & $0.1448(11)$ & $0.1436(11)$ & $0.1429(11)$ &
    $0.1432(11)$ & $0.1418(11)$ \\
    $C_2$ & $32$ & $0.1432(06)$ & $0.1430(06)$ & $0.1429(06)$ &
    $0.1429(06)$ & $0.1427(06)$ \\
    \hline
  \end{tabular*}
  \caption{Comparison of different finite size correction formulae of
    $\mps$ applied to ensembles where we have different volumes
    available, while all other parameters stay fixed. $a\mps(L)$ is
    the measured value, $a\mps^\mathrm{GL}(\infty)$ the infinite 
    volume limit value from GL, $a\mps^\mathrm{CDH,6}(\infty)$ 
    the infinite volume limit value from CDH, taking $I_m^{(6)}$ into account 
    and $a\mps^\mathrm{CDH}(\infty)$ and   
    $a\mps^\mathrm{CDH^{m}}(\infty)$   
    the infinite volume limit value from CDH, without $I_m^{(6)}$. In the last 
    column the CDH formulae with a re-expansion in the quark mass is used.
    }
  \label{tab:Rmps}
\end{table}

\begin{table}[t!]
  \centering
  \begin{tabular*}{1.\textwidth}{@{\extracolsep{\fill}}lccccc}
    \hline\hline
    $\Bigl.\Bigr.$ & $L/a$ & $a\fps(L)$ & $a\fps^\mathrm{GL}(\infty)$ & 
    $a\fps^\mathrm{CDH}(\infty)$ & $a\fps^\mathrm{CDH^{m}}(\infty)$\\
    \hline\hline
    $B_1$ & $24$ & $0.0646(4)$ & $0.0663(4)$ & $0.0662(4)$ &
    $0.0663(4)$ \\
    $B_6$ & $32$ & $0.0663(2)$ & $0.0666(2)$ & $0.0665(2)$ &
    $0.0665(2)$ \\
    \hline
    $C_6$ & $20$ & $0.0508(5)$ & $0.0546(5)$ & $0.0551(5)$ &
    $0.0558(5)$ \\
    $C_5$ & $24$ & $0.0558(5)$ & $0.0577(5)$ & $0.0574(5)$ &
    $0.0578(5)$ \\
    $C_2$ & $32$ & $0.0569(2)$ & $0.0573(2)$ & $0.0572(2)$ &
    $0.0572(2)$ \\
    \hline
  \end{tabular*}
  \caption{same as table~\ref{tab:Rmps} but for $\fps$.}
  \label{tab:Rfps}
\end{table}

We can now apply the different formulae to correct $\mps$ and
$\fps$ for finite size effects for ensembles where we have different
volumes available. The result can be found in tables~\ref{tab:Rmps}
and \ref{tab:Rfps}.
In the tables we show the corrected values at infinite volume 
from the different ways of treating finite size effects as 
indicated in the caption. To this end, we take the measured data at a
given linear extent $L$ and provide the corresponding correction to
infinite volume. For larger volumes these extrapolations agree better
and finally converge within the errors. By checking the stability of the
values extrapolated to infinite volume from the different analytical
formulae we can control the applicability of these corrections for a
given lattice size.

All finite size correction formulae provide an
appropriate framework to describe the observed finite volume
effects. However, for ensembles $B_1$ and $C_6$, corresponding to the
smallest lattice extents at these two $\beta$ values, GL seems to
underestimate FSE in $\mps$, an observation which was also made in
refs.~\cite{Orth:2005kq,Giusti:2007hk}. For those cases the resummed
L{\"u}scher formula provides a better description. In the chiral fits
presented in section~\ref{sec:comb}, however, we observe significantly
larger values of $\chi^2/\mathrm{d.o.f}$ when GL is used, and hence we
use CDH and CDH$^m$ in these fits.

In conclusion, these results make us confident that our simulations
have eventually reached a regime of 
pseudo scalar masses and lattice volumes where $\chi$PT formulae can
be used to estimate FSE for $m_\mathrm{PS}$ and $f_\mathrm{PS}$ when
ensembles with $m_\mathrm{PS}L\geqslant 3$ are used. 
In particular, this allows us to control the finite size effects
for these mesonic quantities from our simulations. But it is clear
that in particular the CDH formula is affected by 
large uncertainties, mainly stemming from the poorly known low
energy constants, which are needed as input. Changing their values in
the range suggested in ref.~\cite{Colangelo:2005gd}, however, changes
the estimated finite size effects maximally at the order of about
$20\%$ (of the corrections themselves)~\cite{Boucaud:2008xu}.

Note that in the above discussion we have only used \emph{continuum}
formulae to describe the FSE of our lattice data. However, we believe
that this can be justified by the smallness of the lattice artefacts
discussed above.  In addition, we restate that at this stage we are
only interested in an effective but good description of the FSE since
our goal was to bring the simulation data from different lattice
spacings to a common finite volume as needed to perform the scaling
analysis. In all the fits performed later on, the finite size
corrections are applied to the continuum results.


\section{Combined Fits}
\label{sec:comb}

The main goal of this section is to confront the 
data for $\fps$ and $\mps$ to $\chi$PT and eventually determine low
energy constants (LEC's) of the chiral effective Lagrangian. 

We shall proceed by presenting results from fits including a
combined continuum, infinite-volume and chiral extrapolation of $\mps$ and $\fps$ for 
two $\beta$-values, $\beta=3.9$ and $\beta=4.05$. Our analysis 
will extend the results given in
refs.~\cite{Urbach:2007rt,Dimopoulos:2007qy} by incorporating data for
the renormalisation 
constant $Z_\mathrm{P}$ and the Sommer parameter $r_0/a$ in the
fit. Details on the computation of $Z_\mathrm{P}$ and $r_0/a$ can be
found in refs.~\cite{Dimopoulos:2007fn,Boucaud:2008xu}. We shall include
the correlation matrix of $f_\mathrm{PS}$ and 
$m_\mathrm{PS}$ into the fit (the correlations with $r_0/a$ and
$Z_\mathrm{P}$ are negligible) and also provide a detailed
account of systematic uncertainties. 

\subsection{General Fit Formulae}
\label{sec:fitform}

The formulae we are going to use to describe the chiral and continuum
behaviour of $\fps$, $\mps$ in infinite volume read:
\begin{equation}
  \label{eq:fmps}
  \begin{split}
    (r_0^\chi \mps)^2 &= (r_0^\chi)^2\ \chi_\mu \Biggl[
    1+\xi\log\left(\frac{\chi_\mu}{\Lambda_3^2}\right)+
    D_{m_\mathrm{PS}}(a/r_0^\chi)^2 + T_m^\mathrm{NNLO}\Biggr]\ ,\\
    r_0^\chi\fps &= r_0^\chi
    f_0\Biggl[1-2\xi\log\left(\frac{\chi_\mu}{\Lambda_4^2}\right) +
    D_{f_\mathrm{PS}}(a/r_0^\chi)^2 + T_f^\mathrm{NNLO}\Biggr]\ , \\
  \end{split}
\end{equation}
where $T_{m,f}^\mathrm{NNLO}$ denote the continuum NNLO $\chi$PT terms
\begin{equation}
  \label{eq:nnlo}
  \begin{split}
    T_m^\mathrm{NNLO} &=
    \ \frac{17}{2}\xi^2\left[\log\frac{\chi_\mu}{\Lambda_M^2}\right]^2 +
    4 \xi^2k_M \ ,\\
    T_f^\mathrm{NNLO} &=
    \,-5\xi^2\left[\log\frac{\chi_\mu}{\Lambda_F^2}\right]^2 +
    4\xi^2k_F \ ,\\
  \end{split}
\end{equation}
with 
\begin{equation}
  \label{eq:nnloterms}
  \begin{split}
    \log{ \frac{\Lambda_M^2}{\chi_\mu} } &= \frac{1}{51} \left( 
      28 \log{ \frac{\Lambda_1^2}{\chi_\mu} } + 
      32 \log{ \frac{\Lambda_2^2}{\chi_\mu} } -
      9 \log{ \frac{\Lambda_3^2}{\chi_\mu} } + 
      49 \right) \ ,\\
    \log{ \frac{\Lambda_F^2}{\chi_\mu} } &= \frac{1}{30} \left( 
      14 \log{ \frac{\Lambda_1^2}{\chi_\mu} } + 
      16 \log{ \frac{\Lambda_2^2}{\chi_\mu} } +
      6 \log{ \frac{\Lambda_3^2}{\chi_\mu} } -
      6 \log{ \frac{\Lambda_4^2}{\chi_\mu} } + 
      23 \right) \ ,  
  \end{split}
\end{equation}
and we have defined
\begin{equation}
  \label{eq:defs}
  \xi\equiv \chi_\mu/(4 \pi f_0)^2,\quad \chi_\mu\equiv2 B_0\mu_R,\quad
  \mu_R\equiv\mu_q/Z_\mathrm{P}^{\overline{\mathrm{MS}}}(\mu=2\, \mathrm{GeV}) \; .
\end{equation}
We use a normalisation such that $f_0=\sqrt{2}F_0$, i.e. $f_\pi=130.7\,
\mathrm{MeV}$. For the finite size corrections we shall use the CDH
and CDH$^m$ formulae as discussed in sub-section~\ref{sec:fs}.

The mass and decay constant of the charged pion have been studied up to
NLO~\cite{Munster:2003ba,Scorzato:2004da,Sharpe:2004ny} in the context of
twisted mass chiral perturbation theory (tm$\chi$PT). The regime of quark
masses and lattice spacings at which we have performed the simulations is
such that $\mu_R \gtrsim a \Lambda_\mathrm{QCD}^2$. In the associated power
counting, at maximal twist the NLO tm$\chi$PT expressions for the charged
pion mass and decay constant preserve their continuum form. The inclusion
of the terms proportional to $D_{\mps,\fps}$ parametrising the
lattice artifacts in eq.~(\ref{eq:fmps}) represents an effective way
of including sub-leading discretisation effects appearing at NNLO. 
The data shown in fig.~\ref{fig:cont} (page~\pageref{fig:cont}) nicely
support the $\mu_R$ independence of $D_{\fps,\mps}$, which we assume
in our fit ansatz (\ref{eq:fmps}). 

$r_0^\chi$ is the value of the Sommer parameter in the chiral limit
which is determined for every $\beta$-value, within the global fit,
according to the formula 
\begin{equation}
  \label{eq:r0chi}
  (r_0/a) = (r_0^\chi/a)\ \Bigl[1 + C_1\ r_0^\chi \mu_R + C_2\
  (r_0^\chi \mu_R)^2\Bigr]
\end{equation}
with $\beta$ independent, i.e. continuum
parameters $C_1$ and $C_2$. In addition to the expected $\mu_R^2$
dependence, we also include the term linear in $\mu_R$ in
order to account for possible effects of spontaneously broken chiral
symmetry. Possible lattice artifacts of $\mathcal{O}(a^2)$ cannot be
distinguished from the lattice artifacts included for $r_0^\chi\mps$ and
$r_0^\chi\fps$ in eq.~(\ref{eq:fmps}), and are therefore effectively described
by the parameters $D_{\mps,\fps}$, whereas lattice artifacts of order
$a^2\mu_R$ and $a^2\mu_R^2$ appear only at higher order.

In order to work with a massless renormalisation scheme, the values of the
pseudo scalar renormalisation constant $Z_\mathrm{P}$ are determined
in the chiral limit. They were determined using the RI'-MOM scheme at
scale $\mu'=1/a$ and the values can be found in
table~\ref{tab:r0ZP}. For the present fit all renormalisation constants
$Z_\mathrm{P}$ need to be run from scale $\mu'=1/a$ to a common scale,
which we have conventionally chosen to be $\mu=2\, \mathrm{GeV}$. This
running can be performed perturbatively, but the scale $\mu'$ is to be
determined. This task can be performed for a given set of fit parameters
by setting $f_\pi = 130.7\, \mathrm{MeV}$ where the ratio $\mps/\fps$
assumes its physical value. Hence, we include for $Z_\mathrm{P}$
\begin{equation}
  \label{eq:ZPrunning}
  Z_\mathrm{P}^{\overline{\mathrm{MS}}}(\mu=2\, \mathrm{GeV}) =
  \zeta(\mu, \mu')\ Z_\mathrm{P}^\mathrm{RI'}(\mu'=1/a)\ /\ R(\mu=2\
  \mathrm{GeV}). 
\end{equation}
into the global fit. $\zeta(\mu, \mu')$ describes the running of
$Z_\mathrm{P}$ from scale $\mu'=1/a$ to the scale $\mu=2\,
\mathrm{GeV}$ computed in perturbation theory to four loops in
$\alpha_s$ in ref.~\cite{Chetyrkin:1999pq}. We use the value
$\Lambda_{\overline{\mathrm{MS}}}=0.2567\ \mathrm{GeV}$~\cite{DellaMorte:2008ad}. The
perturbative factor $R(\mu=2\,\mathrm{GeV})=0.8178$ matches the RI'-MOM to the
$\overline{\mathrm{MS}}$ scheme and is known to NNNLO. It should be
noted that the scale $\mu'$ enters only logarithmically into $\zeta$
and the small changes in $\mu'$ therefore do make only a very small
difference in the running of $Z_\mathrm{P}$. However, we include
eq.~(\ref{eq:ZPrunning}) into the global fit in order to correctly
account for the statistical uncertainty of
$Z_\mathrm{P}^\mathrm{RI'}(\mu'=1/a)$.

At NLO, i.e. setting $T_{m,f}^\mathrm{NNLO}\equiv0$, and neglecting
finite size corrections for the moment, there are the following free
parameters to be fitted to the data for $a\fps$, $a\mps$, $r_0/a$ and
$Z_\mathrm{P}^\mathrm{RI'}(\mu'=1/a)$: 
\[
r_0^\chi f_0,\ r_0^\chi B_0,\ r_0^\chi\Lambda_{3,4},\ C_{1,2},\
\{r_0^\chi/a\}_\beta,\ \{Z_\mathrm{P}^{\overline{\mathrm{MS}}}(2\
\mathrm{GeV})\}_\beta,\,\
 D_{\mps},\ D_{\fps} ,
\]
where we indicate with the notation $\{...\}_\beta$ that there is one
parameter for each $\beta$-value. When also the NNLO expressions for
$\mps$ and $\fps$ are included there are four more parameters to be
fitted to the data:
\[
r_0^\chi\Lambda_{1,2},\ k_M,\ k_F\ .
\]
As already mentioned, finite size effects are corrected for by using the asymptotic formulae
from CDH and CDH$^m$, which is consistently included in the fit. However,
both additionally depend on the low energy constants $\Lambda_{1,2}$.

It will turn out that we do not have sufficient data to determine all
those parameters from the fit. For this reason we shall add priors for
the NNLO low energy constants where needed. A description
of the procedure to compute the $\chi^2$ can be found in
appendix~\ref{sec:chisqr}. 

As mentioned above, in our analysis, we find that for a given  value of $a\mu_q$ and $\beta$
only the data for $a\mps$ and $a\fps$ are correlated and we use a
correlated $\chi^2$ to account for these. Estimates for the correlation
between $a\fps$ and $a\mps$ can be found in the
tables~\ref{tab:corr3.8}-\ref{tab:corr4.2} in
appendix~\ref{sec:data}. 

\subsection{Analysis strategy}

By in- and excluding certain data points, we build different data-sets. 
For a given data-set we perform the following four different fits:
\begin{enumerate}
\item Fit A: NLO continuum $\chi$PT, $T_{m,f}^\mathrm{NNLO}\equiv0$,
  $D_{\mps,\fps}\equiv0$, priors for $r_0\Lambda_{1,2}$

\item Fit B: NLO continuum $\chi$PT, $T_{m,f}^\mathrm{NNLO}\equiv0$,
  $D_{\mps,\fps}$ fitted, priors for $r_0\Lambda_{1,2}$

\item Fit C: NNLO continuum $\chi$PT, $D_{\mps,\fps}\equiv0$,
  priors for $r_0\Lambda_{1,2}$ and $k_{M,F}$

\item Fit D: NNLO continuum $\chi$PT, $D_{\mps,\fps}$
  fitted, priors for $r_0\Lambda_{1,2}$ and $k_{M,F}$

\end{enumerate}
In the above list of different fits, setting $D_{\mps,\fps}\equiv0$
corresponds to a constant continuum extrapolation of $\fps$ and
$\mps$. All these fits are applied to every data-set summarised in
table~\ref{tab:datasets}, and are repeated with both CDH and CDH$^m$,
respectively. While we include or exclude $\mps$ and $\fps$ data
corresponding to the data-set, for $r_0/a$ we have always used only
the largest volume available, since an effective description of finite
volume effects is not available for this quantity.

These different fits include the systematic uncertainties of lattice
spacing artefacts, finite size effects, higher order contributions in
chiral perturbation theory and the extrapolation of $r_0/a$ to the
chiral limit. Our strategy is to build the distribution of all
possible fits using the available data-sets, weighting with the
corresponding confidence level. This procedure is analogous to what
has been done in ref.~\cite{Durr:2008zz}. In table~\ref{tab:results}
we give our best estimates for the fit parameters following this
procedure. The best estimates and also quantities derived from those
are obtained as the median of the weighted distribution from a total
of 76 different fits. All fits are repeated for 1000 bootstrap samples
which are used to estimate the statistical errors, while the
systematic uncertainties are estimated from the 68\% confidence interval of
the weighted distribution, as explained in
appendix~\ref{sec:fitdetails}. As an example we show in
figure~\ref{fig:asqrfits} the result for fit type B on data-set 1,
which is also further discussed in the appendix.

\begin{figure}[t!]
  \centering
  \subfigure[\label{fig:asqrr0}]%
  {\includegraphics[width=0.46\linewidth]{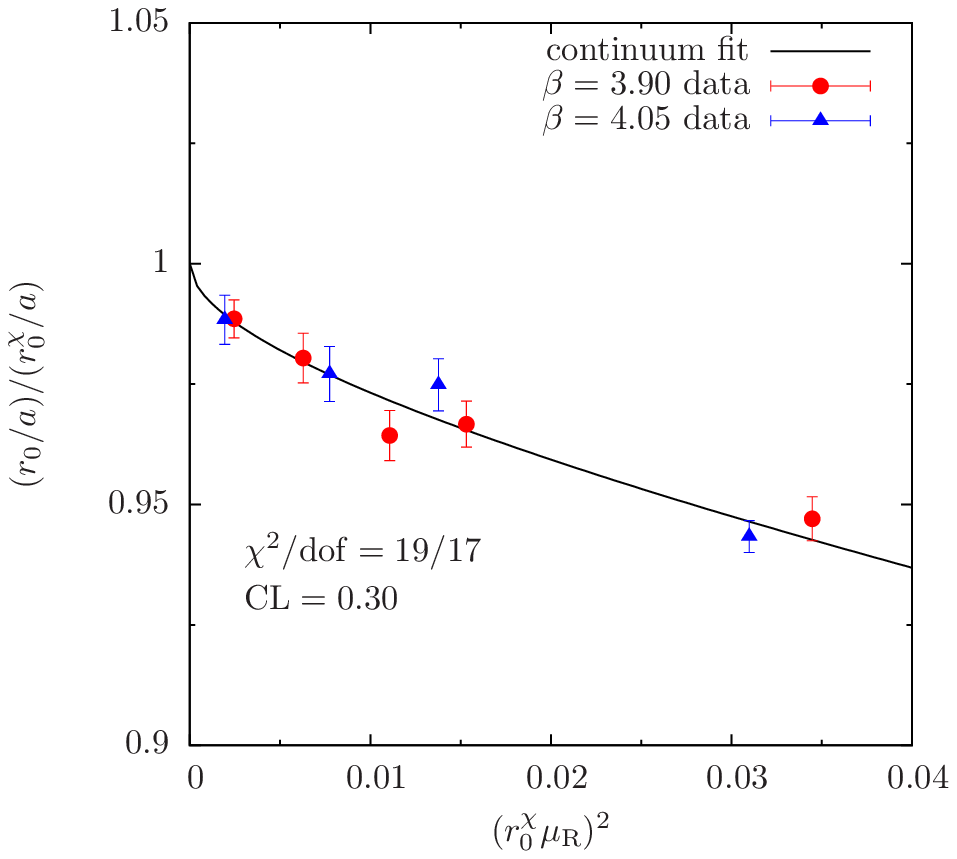}}
  \quad
  \subfigure[\label{fig:asqrf}]%
  {\includegraphics[width=0.44\linewidth]{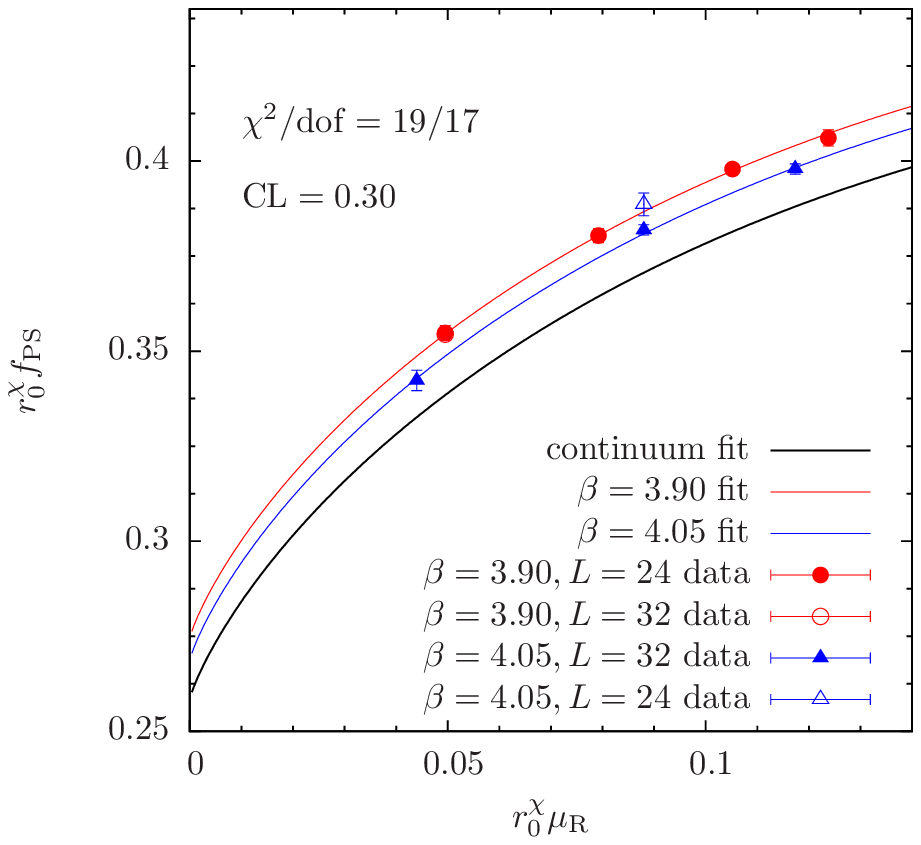}}\\

  \subfigure[\label{fig:asqrmpssq}]%
  {\includegraphics[width=0.45\linewidth]{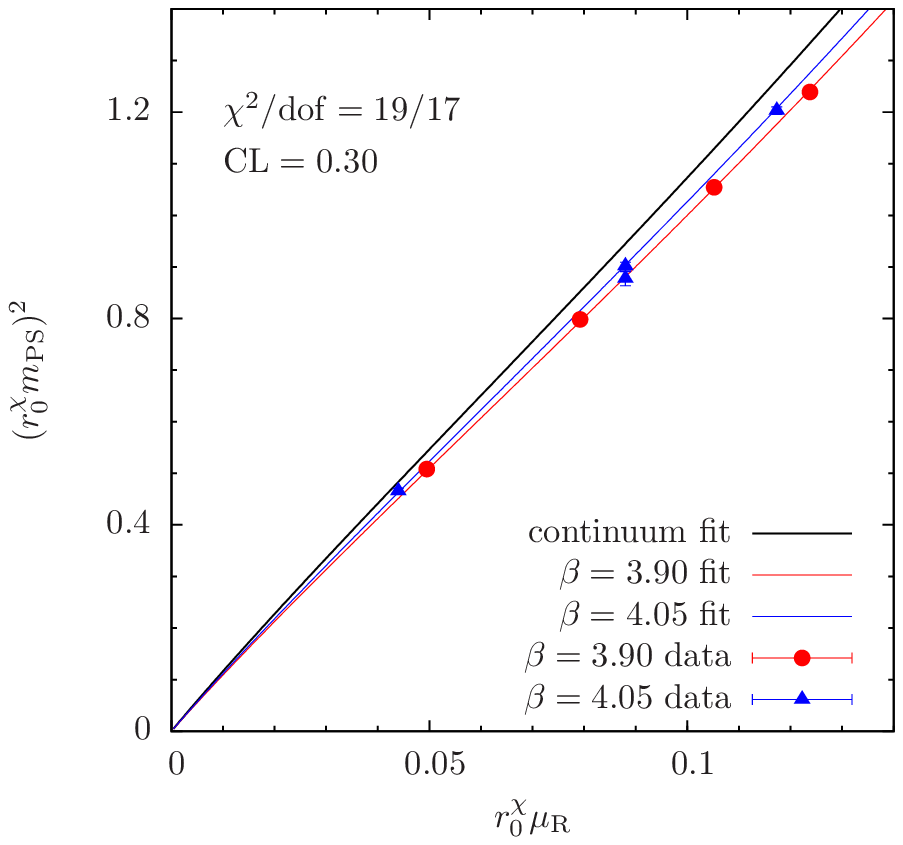}}
  \quad
  \subfigure[\label{fig:asqrmpssqovmu}]%
  {\includegraphics[width=0.44\linewidth]{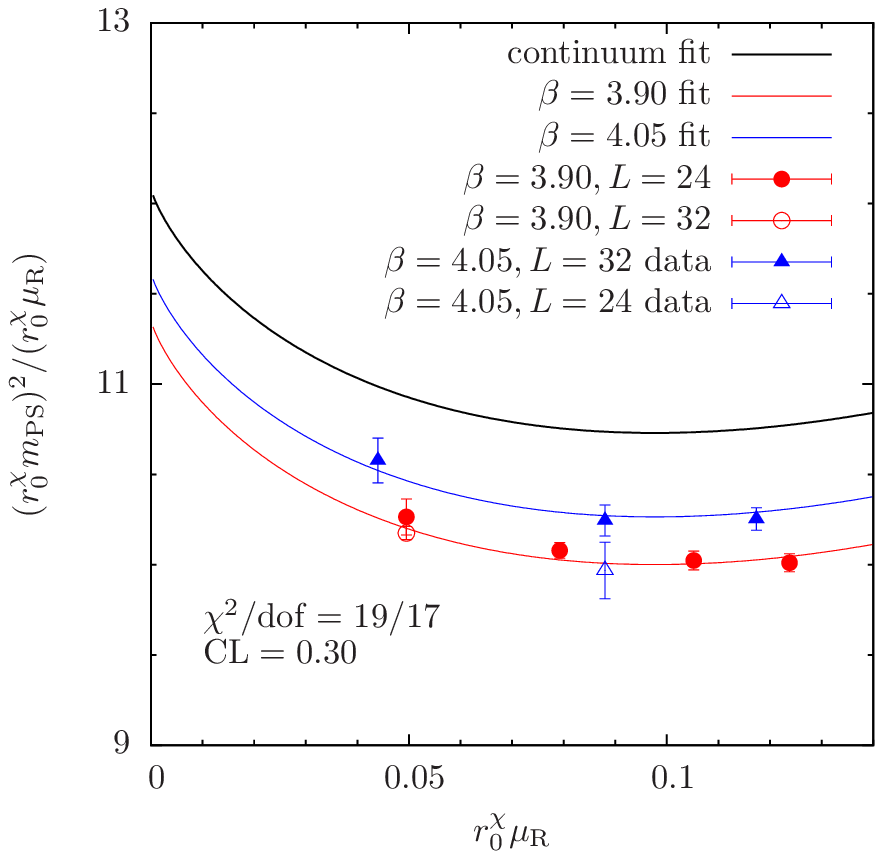}}

  \caption{(a) $(r_0/a)/(r_0^\chi/a)$ as a function of
    $(r^\chi_0\mu_\mathrm{R})^2$. 
    (b) $r^\chi_0\fps$, (c) $(r^\chi_0\mps)^2$ and (d)
    $(r^\chi_0\mps)^2/(r^\chi_0\mu_\mathrm{R})$ as a function of
    $r^\chi_0\mu_\mathrm{R}$. The plots are for fit B on data-set
    1. Circles (triangles) represent data points from $\beta=3.90$
    ($\beta=4.05$). The value of $\chi^2/\mathrm{dof}$ obtained for
    this fit is $19/17$. Note that in these figures we did not propagate the errors of
    $r_0^\chi$ and $Z_\mathrm{P}$. When these errors are included the
    statistical significance of $a^2$-dependence of the data in the figures
    obviously decreases, see table~\protect{\ref{tab:results}}.
  }
  \label{fig:asqrfits}
\end{figure}

We exclude fits from the analysis which we consider as being not
reasonable, i.e. where the NNLO fit gives a worse description
at larger masses than the NLO fit. This is a sign for not having
sufficient data for such a fit. In table~\ref{tab:results} there are
a few quantities which show a strong asymmetry in the estimated
systematics. This uncertainty is coming from the fact, that the
estimate from NLO and NNLO fits for this quantities differs
substantially. However, in the final result the NLO fits have more
weight leading to the observed asymmetries.

In order to obtain further confidence in the uncertainties we quote in
table~\ref{tab:results} we can use the data at $\beta=3.8$ and
$\beta=4.20$. In table~\ref{tab:results2} we compare fits
including these data with the results from table~\ref{tab:results}. It
is clearly visible that the overall uncertainty we quote for all the
quantities comfortably covers the differences among the three
presented fit averages. This fact makes us confident that we
indeed have a good estimate of the systematic uncertainties in our
results. Note that $Z_\mathrm{P}(\beta=4.2)$ is a fit parameter for
the fits including data at $\beta=4.2$. 

We think that this strategy takes into account all systematic effects
that may affect our data and which are accessible to us. This allows
us to provide values for the fit parameters and derived quantities
with controlled statistical and systematic errors. In particular,
since in our fit procedure, finite size corrections are applied in the
{\em continuum}, effects of the neutral pseudo scalar mass on the
finite size effects are absent. The details for our fits can be found
in appendix~\ref{sec:fitdetails}. Note that the main uncertainty to
our results is stemming from our estimates of systematic errors. A
systematic effect we are not able to estimate is the missing strange
and, possibly, charm quark in the sea.

\begin{table}[t!]
  \centering
  \begin{tabular*}{1.\linewidth}{@{\extracolsep{\fill}}lrrrr}
    \hline\hline
    $\Bigl.\Bigr.$ & $\beta=3.8, 3.9, 4.05$ & $\beta=3.9, 4.05$ &
    $\beta=3.9, 4.05, 4.2$ & prior\\
    \hline\hline
    $m_{u,d}\ [\mathrm{MeV}]$                            & $3.84(18)$  & $3.54(26)$    & $3.58(26)$      & - \\   
    $\bar\ell_3$                                         & $3.32(21)$  & $3.50(31)$    & $3.49(9)$       & - \\  
    $\bar\ell_4$                                         & $4.69(17)$  & $4.66(33)$    & $4.63(4)$       & - \\  
    $\bar\ell_1$                                         & $-0.41(58)$ & $-0.33(62)$   & $-0.59(58)$     & $-0.4(6)$ \\ 
    $\bar\ell_2$                                         & $4.31(11)$  & $4.32(11)$    & $4.32(10)$      & $4.3(1)$ \\  
    $f_0\ [\mathrm{MeV}]$                                & $121.7(3)$  & $121.5(1.1)$  & $121.58(8)$     & - \\
    $B_0\ [\mathrm{MeV}]$                                & $2437(120)$ & $2638(200)$   & $2619(190)$     & - \\ 
    $r_0\ [\mathrm{fm}]$                                 & $0.446(9)$  & $0.420(14)$   & $0.429(8)$      & - \\ 
    $C_1$                                                & $-0.09(18)$ & $-0.21(17)$   & $-0.37(14)$     & - \\
    $C_2$                                                & $-1.03(81)$ & $-0.52(77)$   & $0.09(60)$      & - \\
    $\langle r^2\rangle_s^\mathrm{NLO}\ [\mathrm{fm}^2]$ & $0.710(28)$ & $0.715(77)$   & $0.710(9)$      & - \\ 
    $|\Sigma|^{1/3}\ [\mathrm{MeV}]$                     & $262.2(4.0)$& $269.9(6.5)$  & $268.4(6.6)$    & - \\
    $f_\pi/f_0$                                          & $1.0742(81)$& $1.0755(94)$  & $1.0750(8)$     & - \\
    $r^\chi_0/a(\beta=3.80)$                             & $4.462(45)$ & $-$           & $-$             & - \\ 
    $r^\chi_0/a(\beta=3.90)$                             & $5.259(48)$ & $5.316(49)$   & $5.361(39)$     & - \\ 
    $r^\chi_0/a(\beta=4.05)$                             & $6.637(57)$ & $6.661(62)$   & $6.727(48)$     & - \\
    $r^\chi_0/a(\beta=4.20)$                             & $-$         & $$            & $8.358(63)$     & - \\
    $a(\beta=3.80)\ [\mathrm{fm}]$                       & $0.0998(19)$& $-$           & $-$             & - \\
    $a(\beta=3.90)\ [\mathrm{fm}]$                       & $0.0847(15)$& $0.0790(26)$  & $0.0801(14)$    & - \\
    $a(\beta=4.05)\ [\mathrm{fm}]$                       & $0.0672(12)$& $0.0630(20)$  & $0.0638(10)$    & - \\ 
    $a(\beta=4.20)\ [\mathrm{fm}]$                       & $-$         & $-$           & $0.05142(83)$   & - \\
    $Z_\mathrm{P}(\beta=3.80)$                           & $0.431(10)$ & $-$           & $-$             & - \\
    $Z_\mathrm{P}(\beta=3.90)$                           & $0.4390(74)$& $0.4335(84)$  & $0.436(9)$      & - \\
    $Z_\mathrm{P}(\beta=4.05)$                           & $0.455(10)$ & $0.452(13)$   & $0.450(11)$     & - \\
    $Z_\mathrm{P}(\beta=4.20)$                           & $-$         & $-$           & $0.466(19)$     & $0.45(10)$\\
    $D_{\mps}$                                           & $0(1)$      & $-0.7(1.4)$   & $0(2)$          & - \\
    $D_{\fps}$                                           & $0.56(45)$  & $1.68(68)$    & $1.5(6)$        & - \\
    \hline
   \end{tabular*}                                  
  \caption{We compare fit results of fits including data from
    $\beta=3.8$ (first data column) and data from $\beta=4.20$ (third
    data column) with the result where only data from $\beta=3.90$ and
    $\beta=4.05$ have been used (second data column). Note that
    in the case of including data from $\beta=4.20$ the value of
    $Z_\mathrm{P}^{\overline{\mathrm{MS}}}$ is determined from the fit.
    All averages are weighted averages with the confidence levels of
    the individual fits. The errors are statistical and
    systematical, added in quadrature. $B_0$, $\Sigma$ and $m_{u,d}$
    are renormalised in the $\overline{\mathrm{MS}}$ scheme at
    the renormalisation scale $\mu = 2\, \mathrm{GeV}$, as the values
    of $Z_\mathrm{P}$ are in the $\overline{\mathrm{MS}}$ scheme at
    this scale $\mu$. The scale is set by $f_\pi = 130.7\, \mathrm{MeV}$
    as done in ref.~\cite{Boucaud:2007uk}.}
  \label{tab:results2}
\end{table}

Eventually it is interesting to investigate the influence of the way
$r_0/a$ is chirally extrapolated on our final results. In
table~\ref{tab:results}, the systematic effect on $r_0^\chi/a$ is
negligible because of the little variation in the treatment of $r_0/a$
between the different fits. Looking at table~\ref{tab:results2} one
might be surprised that the values for $C_1$ and $C_2$ are both
compatible with zero, even if in figure~\ref{fig:asqrr0} a clear mass
dependence is visible. The reason for this is that with both
parameters the fit is overdetermined and the boostrap samples for
$C_1$ and $C_2$ are strongly anticorrelated ($-0.99$). However, since
we are in this analysis not primarily interested in the estimates for
$C_1$ and $C_2$, but more in the correct estimate of systematics for
other quantities, we keep both parameters in the fits.

In order to address this systematic effect, we have repeated all the
analysis using
\begin{equation}
  \label{eq:r0quad}
  (r_0/a) = (r_0^\chi/a)\ \Bigl[1 + C_2\ (r_0^\chi \mu_R)^2\Bigr]\ ,
\end{equation}
i.e. setting $C_1=0$, as compared to eq.~(\ref{eq:r0chi}). In
table~\ref{tab:results3} we compare the results obtained with
eq.~(\ref{eq:r0quad}) with the ones presented in
table~\ref{tab:results}. The comparison leads to the conclusion that
only the values of $r_0^\chi/a$ are significantly changed, being also 
statistically more precise when eq.~(\ref{eq:r0quad}) is used. Thus, the
systematic uncertainty stemming from the $r_0/a$ extrapolation to the
chiral limit is negligible for physical quantities.

\begin{table}[t!]
  \centering
  \begin{tabular*}{1.\linewidth}{@{\extracolsep{\fill}}lrr}
    \hline\hline
    $\Bigl.\Bigr.$ & table~\protect{\ref{tab:results}} & quadratic \\
    \hline\hline
    $m_{u,d}\ [\mathrm{MeV}]$                            & $3.54(26)$   & $3.56(29)$  \\  
    $\bar\ell_3$                                         & $3.50(31)$   & $3.52(48)$  \\  
    $\bar\ell_4$                                         & $4.66(33)$   & $4.67(25)$  \\  
    $\bar\ell_1$                                         & $-0.33(62)$  & $-0.50(68)$ \\ 
    $\bar\ell_2$                                         & $4.32(11)$   & $4.33(11)$  \\  
    $f_0\ [\mathrm{MeV}]$                                & $121.5(1.1)$ & $121.48(89)$\\
    $B_0\ [\mathrm{MeV}]$                                & $2638(200)$  & $2618(200)$ \\ 
    $r_0\ [\mathrm{fm}]$                                 & $0.420(14)$  & $0.416(14)$ \\ 
    $C_1$                                                & $-0.21(17)$  & $-$         \\
    $C_2$                                                & $-0.52(77)$  & $-1.45(14)$ \\
    $\langle r^2\rangle_s^\mathrm{NLO}\ [\mathrm{fm}^2]$ & $0.715(77)$  & $0.720(60)$ \\ 
    $|\Sigma|^{1/3}\ [\mathrm{MeV}]$                     & $269.9(6.5)$ & $268.8(6.2)$\\
    $f_\pi/f_0$                                          & $1.0755(94)$ & $1.0759(80)$\\
    $r^\chi_0/a(\beta=3.90)$                             & $5.316(49)$  & $5.261(14)$ \\ 
    $r^\chi_0/a(\beta=4.05)$                             & $6.661(62)$  & $6.595(23)$ \\ 
    $a(\beta=3.90)\ [\mathrm{fm}]$                       & $0.0790(26)$ & $0.0790(27)$\\
    $a(\beta=4.05)\ [\mathrm{fm}]$                       & $0.0630(20)$ & $0.0631(21)$\\
    $Z_\mathrm{P}(\beta=3.90)$                           & $0.4335(84)$ & $0.4341(90)$\\
    $Z_\mathrm{P}(\beta=4.05)$                           & $0.452(13)$  & $0.451(13)$ \\ 
    $D_{\mps}$                                           & $-0.7(1.4)$  & $0.1(1.7)$  \\
    $D_{\fps}$                                           & $1.68(68)$   & $2.10(87)$  \\
    \hline
  \end{tabular*}
  \caption{We compare the main results of
    table~\protect{\ref{tab:results}}, where the chiral extrapolation
    of $r_0$ was performed according to
    eq.~(\protect{\ref{eq:r0chi}}), with results obtained with the
    purely quadratic extrapolation of $r_0/a$ to the chiral
    limit in eq.~(\protect{\ref{eq:r0quad}}). For further details see table~\protect{\ref{tab:results}}.}
  \label{tab:results3}
\end{table}

In the tables \ref{tab:results}, \ref{tab:results2} and
\ref{tab:results3} we also quote a number of quantities 
that are derived from the basic fit parameters.  
The values for the low energy constants $\bar\ell_i$,  
\begin{equation}
  \bar{\ell}_i = \log\left(\frac{\Lambda_i^2}{(m_\pi^\pm)^2}\right)
\; i=1,2,3,4
\end{equation}
are determined
using the physical value of the charged pion mass $m_\pi$, which is
the usual convention. 
The scalar condensate can be determined from the fit parameters $B_0$
and $f_0$ in the following way:
\begin{equation}
  \Sigma = -\frac{B_0 f_0^2}{2}
\end{equation}
and the scalar radius at next to leading order is given by
\begin{equation}
  \langle r^2\rangle_s^\mathrm{NLO} =\frac{12}{(4\pi f_0)^2}(\bar{\ell}_4 - 13/12)\, .
\end{equation}


\section{Summary and Outlook}
\label{sec:summ}

In this paper we have demonstrated that Wilson twisted mass fermions,
when tuned to maximal twist, show the expected linear scaling in $a^2$
with an almost negligible coefficient for both the pseudo scalar decay
constant $f_\mathrm{PS}$ and mass $m_\mathrm{PS}$ (see
fig.~\ref{fig:cont} on page~\pageref{fig:cont}). A very similar
behaviour is seen for the nucleon mass
\cite{Alexandrou:2008tn,Alexandrou:2009qu}. It is important to remark
that tuning to maximal
twist has been achieved in our work by demanding that the renormalised
ratio $Z_\mathrm{A} m_\mathrm{PCAC}/\mu_q^\mathrm{min}$ of the PCAC
mass $m_\mathrm{PCAC}$ over the minimal twisted mass
$\mu_q^\mathrm{min}$ employed in our simulations satisfies
$|Z_\mathrm{A} m_\mathrm{PCAC}/\mu_q^\mathrm{min}| < 0.1$. The same
condition is used for the corresponding error of this ratio.

The precise data for $f_\mathrm{PS}$ and $m_\mathrm{PS}$ we have
obtained at several values of the lattice spacing, quark masses and
volumes allowed for a detailed analysis of systematic effects
originating from having to perform a continuum, infinite volume and
chiral limit. In particular, in this way it became possible to
confront our data to analytical predictions from chiral perturbation
theory and to extract a number of low energy constants and derived
quantities as listed in table~\ref{tab:results},
page~\pageref{tab:results}. The main physical 
results we obtain from this analysis are the light quark mass
$m_{u,d}^{\overline{\mathrm{MS}}}(\mu=2\,\mathrm{GeV})=3.54(26)\,\mathrm{MeV}$,
the pseudo scalar decay constant in the chiral limit
$f_0=122(1)\,\mathrm{MeV}$, the scalar condensate
$[\Sigma^{\overline{\mathrm{MS}}}(\mu=2\,\mathrm{GeV})]^{1/3}=270(7)\,\mathrm{MeV}$
and $f_\pi/f_0=1.0755(94)$.

By performing $\mathcal{O}(80)$ different fits, taking into account or
leaving out terms describing e.g. lattice artefacts or NNLO chiral
perturbation theory and including or excluding data sets, we could
achieve a reliable estimate of systematic errors. The final numbers
for the fit parameters and their errors were then derived from the
median, the bootstrap samples and the 68\% confidence level of the
(weighted) distribution of fit parameters originating from the
different fits. Some quantities such as $\bar{l}_3=3.50(31)$,
$\bar{l}_4=4.66(33)$ or the scalar condensate belong to the most
precise determinations available today. For a comparison to other
lattice results we refer the reader to
ref.~\cite{Necco:2009cq,Aoki:2009zb,Scholz:2009yz}.

As a particular effect appearing in the twisted mass formulation of 
lattice QCD, we determined the size of isospin violation in various 
quantities. In this analysis we found, in accordance with theoretical
expectations \cite{Frezzotti:2007qv}, that only the neutral pseudo
scalar mass is significantly affected by the isospin breaking. 

We conclude that Wilson twisted mass fermions at maximal twist provide 
a lattice QCD formulation allowing for precise computations of quantities
in the light meson sector of QCD.  
Clearly, the next step is now to include dynamical strange and charm
degrees of freedom in our simulations and work in this direction is in 
progress \cite{Baron:2008xa,Baron:2009}.

\subsubsection*{Acknowledgments}

The  computer time for this project was made available to us by the
John von Neumann-Institute for Computing on the JUMP and Jugene systems
in J\"ulich and apeNEXT system in Zeuthen, by UKQCD
on the QCDOC machine at Edinburgh, by INFN on the apeNEXT systems in Rome, 
by BSC on Mare-Nostrum in Barcelona (www.bsc.es), 
by the Leibniz Computer centre in Munich on the Altix system and 
by the computer resources
made available by CNRS on the BlueGene system at GENCI-IDRIS Grant
2009-052271 and CCIN2P3 in Lyon.
We
thank these computer centres and their staff for all technical  advice
and help. 
On QCDOC we have made use of Chroma~\cite{Edwards:2004sx} and BAGEL~\cite{BAGEL}
software and we thank members of UKQCD for assistance. For the analysis we used
among others the R language for statistical computing \cite{R:2005}.

This work has been supported in part by  the DFG 
Sonder\-for\-schungs\-be\-reich/ Trans\-regio SFB/TR9-03  and the EU Integrated
Infrastructure Initiative Hadron Physics (I3HP) under contract
RII3-CT-2004-506078.  We also thank the DEISA Consortium (co-funded by
the EU, FP6 project 508830), for support within the DEISA Extreme
Computing Initiative (www.deisa.org).  G.C.R. and R.F. thank MIUR (Italy)
for partial financial support under  the contract PRIN04. 
V.G. and D.P. thank MEC (Spain) for partial financial support under grant 
FPA2005-00711.

\addappheadtotoc
\appendixpage
\appendix

\section{Details of fits and discussion of systematics}
\label{sec:fitdetails}

In this appendix we discuss the details of our data analysis. For a
more comprehensive discussion see ref.~\cite{Amsler:2008zzb}.
For every fit we compute besides the usual $\chi^2$ value
also the associated confidence levels (CL) (sometimes called 
goodness of the fit)
\[
\mathrm{CL}(q=\chi^2,n=\mathrm{dof}) = 1 - P(q/2,\ n/2)\, ,
\] 
where $P$ is the incomplete Gamma function which corresponds 
to the cumulative $\chi^2$-distribution function for the given number of degrees
of freedom $F_{\chi^2}(q,n)$. $F_{\chi^2}(q,n)$ itself 
represents (in the frequentists approach to statistics) the
probability of finding a 
value of $\chi^2\leq q$. A confidence level of $\mathrm{CL}(q)=x$
indicates then the fraction of times the computed value of $\chi^2$ is
larger than $q$, even if the fitted model is correct.

Since we follow a Bayesian approach by including prior knowledge for
some parameters into our fits, the interpretation of $\mathrm{CL}$ is
more a Bayesian credibility level. Hence, we are not going to
determine the most probably correct model on the basis of the values
of the CL's, but we shall rather incorporate all different fits into
the final result. The CL's of each fit are then used as weights,
specifying the impact of a given fit on the final result.

In more detail, for obtaining the final results we first sort out all
physically un-reasonable fits, as will be described below. Then we generate the CL-weighted
distribution of a given fit parameter or of a derived quantity $\theta$
over all the retained fits. The expectation value of $\theta$ is
estimated as the median of the weighted distribution. By performing
this procedure for every bootstrap sample we determine the statistical
error on our estimate for $\theta$. An estimate for the systematic
uncertainty is provided by $68.4\%$ confidence interval of the
weighted distribution. The estimate for the final error is obtained by
adding the statistical and the systematic error in quadrature. 

These systematic errors cover effects from lattice artefacts, from
different orders in $\chi$PT, finite size effects and fitting range of
the quark masses. We present separately the results including data
generated at $\beta=3.8$ and $\beta=4.2$, because they are on a
different level of accuracy due to insufficient precision in tuning to
maximal twist at $\beta=3.8$ and due to the missing estimate of
$Z_\mathrm{P}^\mathrm{RI}$ at $\beta=4.2$. The results for these
averages can be found in table~\ref{tab:results2}, in- and excluding
$\beta=3.8$ and $\beta=4.2$, compared to the result using only
$\beta=3.9$ and $\beta=4.05$ data. The total residual error we quote
for our \emph{final} results covers the difference to the results
including $\beta=3.8$ and $\beta=4.2$ data.

As mentioned before, we vary the data-sets in order to probe the
influence of the fit range and the finite volume effects on the
$\chi$PT fit. The data-sets we use are compiled in
table~\ref{tab:datasets}. We use both CDH and CDH$^m$ to correct for
finite size effects to further check for the influence of those on our
results.

The priors for $r_0\Lambda_{1,2}$ and $k_{M,F}$ are necessary to
obtain stable fits. The priors for $r_0\Lambda_{1,2}$ are taken from
table~\ref{tab:l12} and $k_{M,F}=0\pm 10$ is sufficient.

\begin{table}[t!]
  \centering
  \begin{tabular*}{1.\linewidth}{@{\extracolsep{\fill}}lccccccccccccc}
    \hline\hline
    set & $B_1$ & $B_2$ & $B_3$ & $B_4$ &
    $B_5$ & $B_6$ & $B_7$ & $C_1$ & $C_2$ & $C_3$ & $C_4$ & $C_5$ &
    $C_6$ \\
    \hline\hline
    1   & x & x & x & x &   & x &   & x & x & x &   & x & \\
    2   & x & x & x & x & x & x &   & x & x & x & x & x & \\
    3   & x & x & x & x &   & x & x & x & x & x &   & x & \\
    4   & x & x & x & x & x & x & x & x & x & x & x & x & \\
    5   & x & x & x &   &   & x & x & x & x &   &   & x & \\
    6   &   & x & x & x &   & x &   & x & x & x &   &   & \\
    7   &   & x & x & x & x & x &   & x & x & x & x &   & \\
    8   &   & x & x & x &   & x & x & x & x & x &   &   & \\
    9   & x & x & x & x &   & x &   & x & x & x &   &   & x \\
    10  & x & x & x & x & x & x &   & x & x & x & x &   & x \\
    11  & x & x & x & x &   & x & x & x & x & x &   &   & x \\
    12  & x & x & x & x & x & x & x & x & x & x & x &   & x \\
    13  & x & x & x &   &   & x & x & x & x &   &   &   & x \\
    \hline
  \end{tabular*}\\
  \vspace*{0.3cm}
  \begin{tabular*}{1.\linewidth}{@{\extracolsep{\fill}}lccccccccccccccc}
    set & $A_2$ & $A_3$ & $A_4$ & $B_1$ & $B_2$ & $B_3$ & $B_4$ &
    $B_5$ & $B_6$ & $B_7$ & $C_1$ & $C_2$ & $C_3$ & $C_4$ & $C_5$ \\
    \hline\hline
    14  & x & x &   & x & x & x & x &   & x &   & x & x & x &   & x \\
    15  & x & x & x & x & x & x & x & x & x &   & x & x & x & x & x \\
    16  & x & x &   & x & x & x & x &   & x & x & x & x & x &   & x \\
    17  & x & x & x & x & x & x & x & x & x & x & x & x & x & x & x \\
    18  & x &   &   & x & x & x &   &   & x & x & x & x &   &   & x \\
    19  & x & x &   &   & x & x & x &   & x &   & x & x & x &   &   \\
    20  & x & x & x &   & x & x & x & x & x &   & x & x & x & x &   \\
    21  & x & x &   &   & x & x & x &   & x & x & x & x & x &   &   \\
    \hline
  \end{tabular*}\\
  \vspace*{0.3cm}
  \begin{tabular*}{1.\linewidth}{@{\extracolsep{\fill}}lcccccccccccccc}
    set & $B_1$ & $B_2$ & $B_3$ & $B_4$ &
    $B_5$ & $B_6$ & $B_7$ & $C_1$ & $C_2$ & $C_3$ & $C_4$ & $C_5$ &
    $D_1$ & $D_2$ \\
    \hline\hline
    22  & x & x & x & x &   & x &   & x & x & x &   & x & x & x \\
    23  & x & x & x & x & x & x &   & x & x & x & x & x & x & x \\
    24  & x & x & x & x &   & x & x & x & x & x &   & x & x & x \\
    25  & x & x & x & x & x & x & x & x & x & x & x & x & x & x \\
    26  & x & x & x &   &   & x & x & x & x &   &   & x & x & x \\
    27  &   & x & x & x &   & x &   & x & x & x &   &   & x & x \\
    28  &   & x & x & x & x & x &   & x & x & x & x &   & x & x \\
    29  &   & x & x & x &   & x & x & x & x & x &   &   & x & x \\
    \hline
  \end{tabular*}
  \caption{We list the various sets of ensembles used for the fits. We refer to 
           table~\ref{tab:setup} for the specification of the ensembles 
           used.}
  \label{tab:datasets}
\end{table}  

\subsubsection*{NNLO Fits}

\begin{figure}[t!]
  \centering
  \subfigure[\label{fig:nnlof}]%
  {\includegraphics[width=0.45\linewidth]{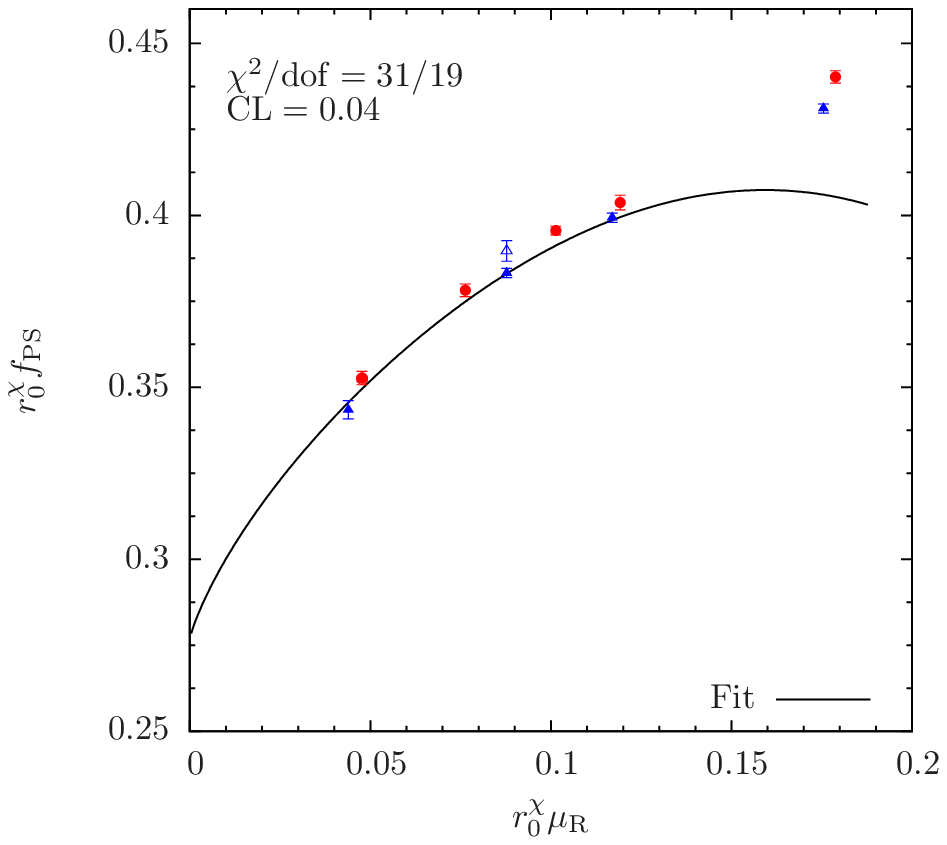}}
  \quad
  \subfigure[\label{fig:nnloforder}]%
  {\includegraphics[width=0.45\linewidth]{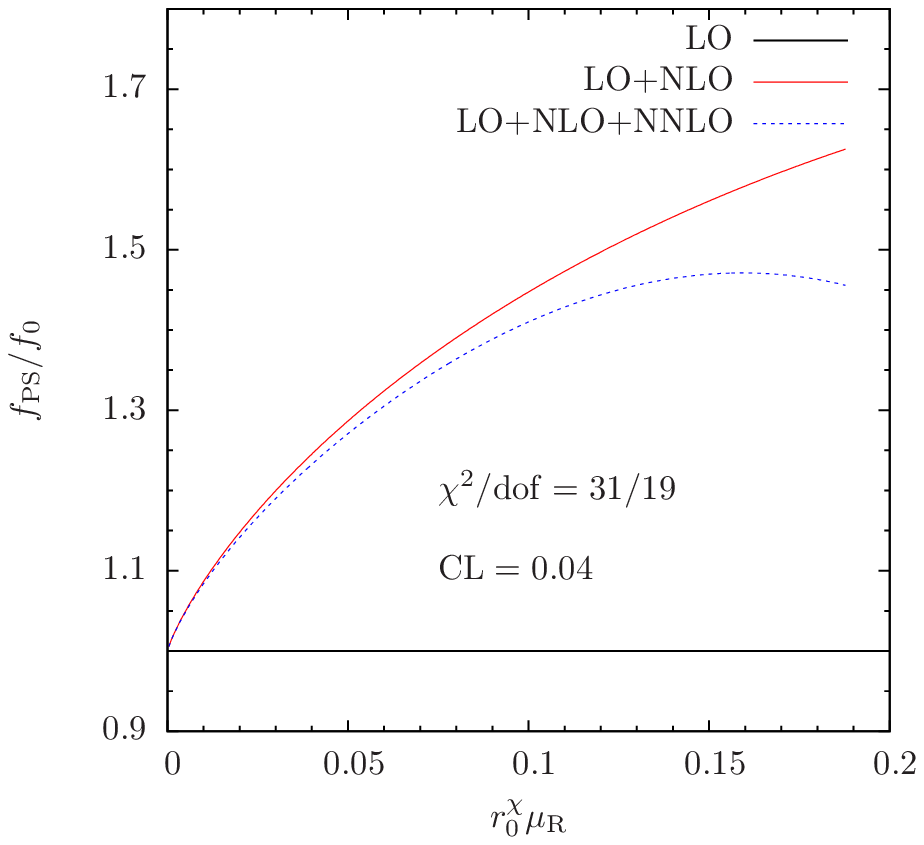}}
  \caption{We show fit C for data set 1, i.e. including NNLO terms,
    but excluding lattice artifacts. This fit was not included in the
    weighted average. In (a) the best fit function for $\fps$ as a
    function of the renormalised quark mass. In (b) we plot $\fps/f_0$
    in LO, NLO and NNLO for comparison.  }
  \label{fig:nnlofits}
\end{figure}

The NNLO $\chi$PT formulae provide in general a better $\chi^2$/d.o.f.
than the corresponding NLO formulae. However, it seems that the larger
masses with $r_0^\chi \mu_\mathrm{R} \gtrsim 0.14$ are necessary to
stabilise these fits. If these masses are left out from the fits then
the NNLO fit curves strongly at large values of
$r_0^\chi\mu_\mathrm{R}$ and actually undershoots the data at larger
masses. This can be seen in figure~\ref{fig:nnlofits}, where we show
in the left panel the data for $r_0\fps$ as a function of
$r_0^\chi\mu_\mathrm{R}$ and the best fit function for fit C. Note
that the data at large values of $r_0^\chi\mu_\mathrm{R}$ are not
included in the fit. In the right panel we show the LO, NLO and the
complete NNLO function for comparison. Evidently, for larger masses,
there is a significant difference between the NLO and NNLO results. It
is in particular suspicious that the NNLO fit shows a stronger
curvature than the NLO fit and that the NNLO fit deviates more from
the data points at large masses -- which are not included in the
fit. On the contrary one would expect the NNLO formulae to describe
the data points at larger masses better than the NLO formulae. On the
other hand, when including the heavier masses, the NNLO fit is able
to describe these data points. To improve the sensitivity of our
lattice data to $\chi$PT at NNLO, additional data points would be needed.

We remark that in ref.~\cite{Frezzotti:2008dr} the inclusion of NNLO
$\chi$PT was found to be necessary in order to obtain an adequate fit to the data
on the pion radius. In ref.~\cite{Frezzotti:2008dr} it was argued that
due to the $\rho$-meson dominance one should expect a slower rate of
convergence of chiral perturbation theory. However, using the data for
$a\fps$ and $a\mps$ alone turns out to be insufficient to obtain the
low energy constants at NNLO. 

We observe this behaviour for all NNLO fits for data sets without the
ensembles at large values of $r_0^\chi\mu_\mathrm{R}\gtrsim 0.14$,
therefore, we do not consider these fits as \emph{reasonable} and
exclude them in the weighted average, even if the CL for this fit
would justify the inclusion. Note however that including them does not
change the final results within the quoted uncertainties.

As already discussed in section~\ref{sec:comb}, one effect of the NNLO
fits included in the final analysis is that the estimated systematic
uncertainty in e.g. $f_0$, $\bar\ell_3$ and $\bar\ell_4$ is
asymmetric. If we analyse NNLO fits only, we would compared to
table~\ref{tab:results} rather obtain $f_0=122.6(1.1)$,
$\bar\ell_3=3.22(66)$, $\bar\ell_4=4.33(34)$. This uncertainty is
included in our final error estimate of those quantities.

\subsubsection*{Lattice Artifacts}

When performing fit B, which uses only NLO $\chi$PT and includes
lattice artifacts, we find that the fitting parameter $D_{\mps}$
parametrising the lattice artifacts in $\mps$ is compatible with zero
within errors. In $\fps$ lattice artifacts are more apparent, even if
the error on $D_{\fps}$ is still large.

However, from the fits we clearly observe that the $\chi^2$ values are
significantly smaller when lattice artifacts are included, and the
weighted distribution is hence dominated by the fits of type B and D. 
As an example for such a fit we show in figure~\ref{fig:asqrfits}
plots for fit B on data-set 1. We show the data together with the
fitted curves for $(r_0/a)/(r_0^\chi/a)$ as a function of
$(r_0^\chi\mu_\mathrm{R})^2$ in (a), for $r_0^\chi\fps$ in (b), for
$(r_0^\chi\mps)^2$ in (c) and for
$(r_0^\chi\mps)^2/(r_0^\chi\mu_\mathrm{R})$ in (d), the last three as
a function of $r_0^\chi\mu_\mathrm{R}$. Note that in
figures~\ref{fig:asqrf}, \ref{fig:asqrmpssq} and
\ref{fig:asqrmpssqovmu} we did not include the error of $r_0^\chi$ and
$Z_\mathrm{P}$ for the data points.

From figure~\ref{fig:asqrfits} one can observe that the fit works very
well, leading to $\chi^2/\mathrm{dof} = 19/17$. In (a) one can see
that the data for $r_0/a$ is also fully compatible with a linear
dependence on $(r_0^\chi\mu_\mathrm{R})^2$ only, as was also discussed
earlier and summarised in table~\ref{tab:results3}.

\subsubsection*{Finite Size Corrections and Effects of Priors}

We have used the CDH and CDH$^m$ formulae for the fits and include
both into the weighted distribution of all fits. It is still
interesting to compare 
two fits, using CDH$^m$ on the one hand and the parameter free GL
formulae on the other hand, while all the rest is identical.  We observe
that in general the $\chi^2$-value  using CDH$^m$ is about a factor of 2
smaller than the one using GL. Therefore, we think that having  extra
parameters (which are included in the fit with priors) is justified.

Another test for uncertainties stemming from FS corrections is to
perform fits where only the largest available volume per $\mu_q$-value
is used. Including
these ensembles (which correspond to data-sets 6,7 and 8) 
tend in general to produce  improved
values of $\chi^2$, while the influence on other fit parameters is
again small. Clearly, it would be desirable to have for all ensembles
values of $\mps L > 4$, where all the FS correction formulae give
very similar results and the uncertainty from finite volumes is hence
much reduced. In this investigation we include this
significant uncertainty in our final error estimate. 

As discussed previously, we have to add priors for $\bar\ell_{1,2}$
and $k_{M,F}$ in order to stabilise the fits. It is therefore
interesting to discuss the influence of the choice for the priors on
the fit results. First of all, the influence of the choice for
$k_{M,F}$ and their errors has negligible effect on our fit results.

In case of $\bar\ell_{1,2}$ the influence depends on whether or not
NNLO terms are included in the chiral expansion. While doubling the
error estimates of $\bar\ell_{1,2}$ or changing their values within
the error range has negligible effects in the case of a NLO chiral fit
(fits A and B), many fit results change drastically in the case of
NNLO chiral fits (fits C and D): we observe in the latter case that
the best fit values e.g. for $\bar\ell_3$ and $\bar\ell_4$ differ
substantially from the NLO fit result and also from the estimates
given in table~\ref{tab:l12}. At the same time the errors become much
larger on those quantities, up to 50\%.

Moreover, the $\chi^2$ minimisation process becomes more difficult, the
$\chi^2$ functions shows many nearby local minima in parameter
space. These facts confirm once more that $\fps$ and $\mps$ data are
not sensitive to NNLO terms in the chiral expansion.

\subsubsection*{Chiral Extrapolation of $Z_\mathrm{P}^\mathrm{RI'}$}

The chiral extrapolation of $Z_\mathrm{P}^\mathrm{RI'}$ is not
performed in the fits, but the chiral value plus error estimates is
input to the fit. In order to check whether this might lead the fits
into a wrong direction, we have performed fits with doubled
uncertainty on the values of $Z_\mathrm{P}^\mathrm{RI'}$. It turns out
that this has negligible effect on our results. 


\section[Chi-square Implementation]{$\chi^2$ Implementation}
\label{sec:chisqr}

We are attempting to fit partially correlated data including
priors for some of the parameters. With our method we closely
follow the approach discussed in ref.~\cite{Lepage:2001ym}. 

The data for $\fps$, $\mps$, $r_0/a$ and $Z_\mathrm{P}^\mathrm{RI'}$ at the
various values of the lattice spacing and the different volumes is 
described 
by the following set of fit parameters
\[
r_0^\chi f_0,\ r_0^\chi B_0,\ r_0^\chi\Lambda_{1-4},\ k_{M,F},\ C_{1,2},\
\{r_0^\chi/a\}_\beta,\ \{Z_\mathrm{P}^{\overline{\mathrm{MS}}}(2\
\mathrm{GeV})\}_\beta,\,\
 D_{\mps},\ D_{\fps} .
\]
With the notation $\{...\}_\beta$ we mean that there is one fit parameter
for each lattice spacing involved in 
the fit. Those parameters can be used, employing eqs.~(\ref{eq:fmps},
\ref{eq:r0chi}) and (\ref{eq:ZPrunning}), 
to compute predictions for $a\mps$, $a\fps$, $r_0/a$
and $Z_\mathrm{P}^{\overline{\mathrm{MS}}}$. We shall denote these predictions by 
\[
\mathcal{P}_{\mps}, \mathcal{P}_{\fps},
\mathcal{P}_{r_0}, \mathcal{P}_{Z_\mathrm{P}}
\]
and they all depend on a sub-list of fit parameters and on
the spatial extend $L/a$, as presented in section~\ref{sec:fitform}. 

We adopt the following definition for the $\chi^2$-function at fixed
value of $\beta$:
\begin{equation}
  \label{eq:chisq}
  \begin{split}
    \chi^2(\beta) =& \sum_{a\mu_q}
    \begin{pmatrix}
      \mathcal{P}_{\mps}-a\mps\\
      \mathcal{P}_{\fps}-a\fps\\
    \end{pmatrix}^T
    C^{-1}
    \begin{pmatrix}
      \mathcal{P}_{\mps}-a\mps\\
      \mathcal{P}_{\fps}-a\fps\\
    \end{pmatrix} + \\ 
    &+ \sum_{a\mu_q}\left(\frac{\mathcal{P}_{r_0}-r_0/a}{\delta[r_0/a]}\right)^2 +
    \left(\frac{\mathcal{P}_{Z_\mathrm{P}}-Z_\mathrm{P}}{\delta[Z_\mathrm{P}]}\right)^2\ .
  \end{split}
\end{equation}
The matrix $C$ is the properly normalised covariance matrix, see
e.g. refs.~\cite{Michael:1993yj,Michael:1994sz}. 
The last two terms in eq.~(\ref{eq:chisq}) are included in order to
appropriately include the measurements for $r_0/a$ and $Z_\mathrm{P}^\mathrm{RI'}$. 
Values for the correlation among $\mps$ and $\fps$ are given
in the tables~\ref{tab:corr3.8}-\ref{tab:corr4.2}. The full $\chi^2$
is obtained by summing $\chi^2(\beta)$ over all $\beta$-values. 

In case we need to add prior knowledge for a fit parameter $\theta$ in
order to stabilise the fit we include additional terms of the form
\begin{equation}
  \label{eq:bayes}
  \left(\frac{\mathcal{P}_{\theta}-\theta}{\delta[\theta]}\right)^2
\end{equation}
to the fit, assuming Gaussian error distribution for $\theta$ with
standard deviation $\delta[\theta]$.
Here the predicted values $\mathcal{P}_{\theta}$ are the fit
parameters and the ``data'' with errors are the priors. The term
eq.~(\ref{eq:bayes}) follows from Bayesian statistics, see
ref.~\cite{Lepage:2001ym} and references therein. 

In detail the computation of $\chi^2$ is performed in the following
steps for fixed values of the fit parameters:
\begin{enumerate}
\item \emph{compute infinite volume and continuum predictions}:\\
  compute the \emph{continuum} predictions for $\mps$ and $\fps$ using
  the formulae~(\ref{eq:fmps}) in infinite volume for given values of
  $r_0^\chi\mu_R$. Determine $r_0^\chi/a$ for each $\beta$-value
  according to eq.~(\ref{eq:r0chi}) and use
  it to estimate the scales $1/a$ from $f_\pi$.

\item \emph{compute finite volume predictions in the continuum}:\\
  now apply finite size corrections to scale the data to the finite
  volumes used in the lattice simulations.

\item \emph{compute finite $a$ predictions}:\\
  use $r_0^\chi/a$ to compute the estimates of $a\mps$ and $a\fps$ for
  finite values of the lattice spacing $a$, depending on the
  coefficients $D_{\mps,\fps}$.

\item \emph{compute $\chi^2$}:\\
  the finite volume and finite $a$ predictions for $\mps$ and $\fps$
  can now be compared to the data and the $\chi^2$ can be computed as
  detailed above. The $\chi^2$ includes also the corresponding terms
  for the chiral extrapolation of $r_0/a$ and for the estimate of
  $Z_\mathrm{P}^{\overline{\mathrm{MS}}}$.
\end{enumerate}
The errors are estimated using the bootstrap method. For $\mps$ and
$\fps$ we use bootstrap samples as generated from the raw data. For
$r_0/a$ and $Z_\mathrm{P}^\mathrm{RI'}$ those are generated from a
Gaussian distribution with mean and standard-deviation properly
chosen. This is justified since we checked that the correlation of
$\mps$ and $\fps$ to $r_0/a$ and $Z_\mathrm{P}^\mathrm{RI'}$ is
negligible. This procedure allows to estimate errors on primary
quantities, such as fit parameters for instance, as well as on derived
quantities, like e.g. the lattice spacing.


\clearpage
\section{Data Tables}
\label{sec:data}


\begin{table}[h!]
  \centering
  \begin{tabular*}{1.0\textwidth}{@{\extracolsep{\fill}}lcccccc}
    \hline\hline
    & $a\mu_q$ & $am_\mathrm{PS}$ & $af_\mathrm{PS}$ &
    $am_\mathrm{PCAC}$ & $r_0/a$ & $L/a$ \\
    \hline\hline
    $A_1$ & $0.0060$ & $0.1852(9)$ & $0.0770(8)$ & $+0.0019(4)$ & $4.321(32)$ & $24$ \\
    $A_2$ & $0.0080$ & $0.2085(8)$ & $0.0835(4)$ & $+0.0008(3)$ & $4.440(34)$ & $24$ \\
    $A_3$ & $0.0110$ & $0.2424(5)$ & $0.0892(3)$ & $-0.0002(5)$ & $4.362(21)$ & $24$ \\
    $A_4$ & $0.0165$ & $0.2957(5)$ & $0.0969(2)$ & $-0.0017(2)$ & $4.264(14)$ & $24$ \\
    $A_5$ & $0.0060$ & $0.1831(6)$ & $0.0784(4)$ & $+0.0005(4)$ & NA          & $20$ \\
    \hline
  \end{tabular*}
  \caption{Data at $\beta=3.8$}
  \label{tab:data3.8}
\end{table}

\begin{table}[h!]
  \centering
  \begin{tabular*}{1.0\textwidth}{@{\extracolsep{\fill}}lcccccccc}
    \hline\hline
    & $a\mu_q$ & $am_\mathrm{PS}$ & $af_\mathrm{PS}$ &
    $am_\mathrm{PCAC}$ & $r_0/a$ & $L/a$ \\
    \hline\hline
    $B_1$ & $0.0040$ & $0.1362(7)$ & $0.0646(4)$ & $+0.00017(17)$ & $5.196(28)$ & $24$ \\
    $B_2$ & $0.0064$ & $0.1694(4)$ & $0.0705(4)$ & $-0.00009(17)$ & $5.216(27)$ & $24$ \\
    $B_3$ & $0.0085$ & $0.1940(5)$ & $0.0742(2)$ & $-0.00052(17)$ & $5.130(28)$ & $24$ \\
    $B_4$ & $0.0100$ & $0.2100(5)$ & $0.0759(4)$ & $-0.00097(26)$ & $5.143(25)$ & $24$ \\
    $B_5$ & $0.0150$ & $0.2586(7)$ & $0.0830(3)$ & $-0.00145(42)$ & $5.039(24)$ & $24$ \\
    $B_6$ & $0.0040$ & $0.1338(2)$ & $0.0663(2)$ & $+0.00022(11)$ & $5.259(21)$ & $32$ \\
    $B_7$ & $0.0030$ & $0.1167(4)$ & $0.0633(3)$ & $+0.00030(14)$  & NA & $32$ \\
    \hline
  \end{tabular*}
  \caption{Data at $\beta=3.9$}
  \label{tab:data3.9}
\end{table}

\begin{table}[h!]
  \centering
  \begin{tabular*}{1.0\textwidth}{@{\extracolsep{\fill}}lcccccccc}
    \hline\hline
    & $a\mu_q$ & $am_\mathrm{PS}$ & $af_\mathrm{PS}$ &
    $am_\mathrm{PCAC}$ & $r_0/a$ & $L/a$ \\
    \hline\hline
    $C_1$ & $0.003$ & $0.1038(6) $ & $0.0500(4)  $ & $+0.00036(14)$ & $6.584(34)$ & $32$ \\
    $C_2$ & $0.006$ & $0.1432(6) $ & $0.0569(2)  $ & $-0.00004(14)$ & $6.509(38)$ & $32$ \\
    $C_3$ & $0.008$ & $0.1651(5) $ & $0.0595(2)  $ & $-0.00065(13)$ & $6.494(36)$ & $32$ \\
    $C_4$ & $0.012$ & $0.2025(6) $ & $0.0644(2)  $ & $-0.00092(14)$ & $6.284(22)$ & $32$ \\
    $C_5$ & $0.006$ & $0.1448(11)$ & $0.0558(5)  $ & $-0.00027(19)$ & NA & $24$ \\
    $C_6$ & $0.006$ & $0.1520(15)$ & $0.0508(5)  $ & $+0.00002(20)$ & NA & $20$ \\
    \hline
  \end{tabular*}
  \caption{Data at $\beta=4.05$}
  \label{tab:data4.05}
\end{table}

\begin{table}[h!]
  \centering
  \begin{tabular*}{1.0\textwidth}{@{\extracolsep{\fill}}lcccccccc}
    \hline\hline
    & $a\mu_q$ & $am_\mathrm{PS}$ & $af_\mathrm{PS}$ &
    $am_\mathrm{PCAC}$ & $r_0/a$ & $L/a$ \\
    \hline\hline
    $D_1$ & $0.0020$  & $0.0740(3)$ & $0.0398(2)$ & $+0.00006(6)$ &
    $8.295(45)$ & $48$ \\
    $D_2$ & $0.0065$ & $0.1326(5)$ & $0.0465(3)$ & $-0.00032(11)$&
    $8.008(29)$ & $32$ \\
    \hline
  \end{tabular*}
  \caption{Data at $\beta=4.20$}
  \label{tab:data4.20}  
\end{table}

\begin{table}[h!]
  \centering
  \begin{tabular}[t]{lc}
    \hline\hline
    $\beta$ & $Z_\mathrm{P}^\mathrm{RI'}(\mu'=1/a)$ \\ 
    \hline\hline
    $3.80$ & $0.340(10)$ \\
    $3.90$ & $0.384(08)$ \\
    $4.05$ & $0.417(11)$ \\
    \hline
  \end{tabular}
  \caption{Data for $Z_\mathrm{P}$
    values in the RI'-MOM scheme at scale $\mu'=1/a$.}
  \label{tab:r0ZP}
\end{table}  

\begin{table}[h!]
  \centering
  \begin{tabular*}{.5\textwidth}{@{\extracolsep{\fill}}cccc}
    \hline\hline
    $A_1$ & $A_2$ & $A_3$ & $A_4$ \\
    \hline\hline
    $-0.72$ & $-0.66$ & $-0.03$ & $0.06$ \\
    \hline
  \end{tabular*}
  \caption{Correlation among $a\fps$ and $a\mps$ at $\beta=3.8$}
  \label{tab:corr3.8}
\end{table}

\begin{table}[h!]
  \centering
  \begin{tabular*}{.8\textwidth}{@{\extracolsep{\fill}}ccccccc}
    \hline\hline
    $B_1$ & $B_2$ & $B_3$ & $B_4$ & $B_5$ & $B_6$ & $B_7$ \\
    \hline\hline
    $-0.53$ & $-0.72$ & $-0.44$ & $0.10$ & $0.34$ & $-0.62$ & $-0.74$ \\
    \hline
  \end{tabular*}
  \caption{Correlation among $a\fps$ and $a\mps$ at $\beta=3.9$}
  \label{tab:corr3.9}
\end{table}

\begin{table}[h!]
  \centering
  \begin{tabular*}{.7\textwidth}{@{\extracolsep{\fill}}cccccc}
    \hline\hline
    $C_1$ & $C_2$ & $C_3$ & $C_4$ & $C_5$ & $C_6$ \\
    \hline\hline
    $-0.39$ & $-0.44$ & $-0.34$ & $-0.07$ & $-0.41$ & $-0.44$ \\
    \hline
  \end{tabular*}
  \caption{Correlation among $a\fps$ and $a\mps$ at $\beta=4.05$}
  \label{tab:corr4.05}
\end{table}

\begin{table}[h!]
  \centering
  \begin{tabular*}{.25\textwidth}{@{\extracolsep{\fill}}cc}
    \hline\hline
    $D_1$ & $D_2$ \\
    \hline\hline
    $-0.44$ & $-0.22$ \\
    \hline
  \end{tabular*}
  \caption{Correlation among $a\fps$ and $a\mps$ at $\beta=4.2$}
  \label{tab:corr4.2}
\end{table}

\begin{table}[h!]
  \centering
  \begin{tabular*}{.8\textwidth}{@{\extracolsep{\fill}}lcccc}
    \hline\hline
    & $a\mps^{\pm}$ & $a\mps^0$ & $am_\mathrm{V}^\pm$ & $am_\mathrm{V}^0$\\
    \hline\hline
    $B_1$ & $0.1362(7)$ & $0.109(07)$ & $0.404(22)$ & $0.391(15)$\\
    $B_2$ & $0.1694(4)$ & $0.134(10)$ & $0.422(09)$ & $0.434(19)$ \\
    $B_3$ & $0.1940(5)$ & $0.169(11)$ & $0.428(08)$ & $0.424(14)$\\
    $B_6$ & $0.1338(2)$ & $0.110(08)$ & $0.416(14)$ & $0.409(21)$\\
    \hline
    $C_1$ & $0.1038(6)$ & $0.090(06)$ & $0.335(12)$ & $0.352(23)$\\
    $C_2$ & $0.1432(6)$ & $0.123(06)$ & $0.347(08)$ & $0.344(13)$\\
    \hline
  \end{tabular*}
  \caption{Comparison of $a\mps^{\pm}$ to $a\mps^0$ and
    $m_\mathrm{V}^\pm$ to $am_\mathrm{V}^0$ for the ensembles where 
    the corresponding neutral masses have been computed.}
  \label{tab:mpi0}
\end{table}

\clearpage

\bibliographystyle{h-physrev5}
\bibliography{bibliography}

\begin{thebibliography}{10}

\bibitem{Jansen:2008vs}
K.~Jansen,
\newblock PoS {\bf LATTICE2008}, 010 (2008),
  \href{http://arxiv.org/abs/0810.5634}{{\tt arXiv:0810.5634 [hep-lat]}}.

\bibitem{Scholz:2009yz}
E.~E. Scholz,
\newblock \href{http://arxiv.org/abs/0911.2191}{{\tt arXiv:0911.2191
  [hep-lat]}}.

\bibitem{Frezzotti:2000nk}
{\bf ALPHA} Collaboration, R.~Frezzotti, P.~A. Grassi, S.~Sint and P.~Weisz,
\newblock JHEP {\bf 08}, 058 (2001),
  \href{http://arxiv.org/abs/hep-lat/0101001}{{\tt hep-lat/0101001}}.

\bibitem{Frezzotti:2003ni}
R.~Frezzotti and G.~C. Rossi,
\newblock JHEP {\bf 08}, 007 (2004),
  \href{http://arxiv.org/abs/hep-lat/0306014}{{\tt hep-lat/0306014}}.

\bibitem{Frezzotti:2004wz}
R.~Frezzotti and G.~C. Rossi,
\newblock JHEP {\bf 10}, 070 (2004),
  \href{http://arxiv.org/abs/hep-lat/0407002}{{\tt hep-lat/0407002}}.

\bibitem{Boucaud:2007uk}
{\bf ETM} Collaboration, P.~Boucaud {\em et~al.},
\newblock Phys. Lett. {\bf B650}, 304 (2007),
  \href{http://arxiv.org/abs/hep-lat/0701012}{{\tt arXiv:hep-lat/0701012}}.

\bibitem{Blossier:2007vv}
{\bf ETM} Collaboration, B.~Blossier {\em et~al.},
\newblock JHEP {\bf 04}, 020 (2008), \href{http://arxiv.org/abs/0709.4574}{{\tt
  arXiv:0709.4574 [hep-lat]}}.

\bibitem{Cichy:2008gk}
K.~Cichy, J.~Gonzalez~Lopez, K.~Jansen, A.~Kujawa and A.~Shindler,
\newblock Nucl. Phys. {\bf B800}, 94 (2008),
  \href{http://arxiv.org/abs/0802.3637}{{\tt arXiv:0802.3637 [hep-lat]}}.

\bibitem{Boucaud:2008xu}
{\bf ETM} Collaboration, P.~Boucaud {\em et~al.},
\newblock Comput. Phys. Commun. {\bf 179}, 695 (2008),
  \href{http://arxiv.org/abs/0803.0224}{{\tt arXiv:0803.0224 [hep-lat]}}.

\bibitem{Alexandrou:2008tn}
{\bf ETM} Collaboration, C.~Alexandrou {\em et~al.},
\newblock Phys. Rev. {\bf D78}, 014509 (2008),
  \href{http://arxiv.org/abs/0803.3190}{{\tt arXiv:0803.3190 [hep-lat]}}.

\bibitem{Jansen:2008wv}
{\bf ETM} Collaboration, K.~Jansen, C.~Michael and C.~Urbach,
\newblock Eur. Phys. J. {\bf C58}, 261 (2008),
  \href{http://arxiv.org/abs/0804.3871}{{\tt arXiv:0804.3871 [hep-lat]}}.

\bibitem{Shindler:2007vp}
A.~Shindler,
\newblock Phys. Rept. {\bf 461}, 37 (2008),
  \href{http://arxiv.org/abs/0707.4093}{{\tt arXiv:0707.4093 [hep-lat]}}.

\bibitem{Blossier:2009bx}
{\bf ETM} Collaboration, B.~Blossier {\em et~al.},
\newblock JHEP {\bf 07}, 043 (2009), \href{http://arxiv.org/abs/0904.0954}{{\tt
  arXiv:0904.0954 [hep-lat]}}.

\bibitem{Jansen:2009hr}
{\bf ETM} Collaboration, K.~Jansen, C.~McNeile, C.~Michael and C.~Urbach,
\newblock Phys. Rev. {\bf D80}, 054510 (2009),
  \href{http://arxiv.org/abs/0906.4720}{{\tt arXiv:0906.4720 [hep-lat]}}.

\bibitem{McNeile:2009mx}
{\bf ETM} Collaboration, C.~McNeile, C.~Michael and C.~Urbach,
\newblock Phys. Lett. {\bf B674}, 286 (2009),
  \href{http://arxiv.org/abs/0902.3897}{{\tt arXiv:0902.3897 [hep-lat]}}.

\bibitem{Blossier:2009hg}
B.~Blossier {\em et~al.},
\newblock \href{http://arxiv.org/abs/0909.3187}{{\tt arXiv:0909.3187
  [hep-lat]}}.

\bibitem{Jansen:2009tt}
K.~Jansen and A.~Shindler,
\newblock \href{http://arxiv.org/abs/0911.1931}{{\tt arXiv:0911.1931
  [hep-lat]}}.

\bibitem{Jansen:2003ir}
{\bf \xlf} Collaboration, K.~Jansen, A.~Shindler, C.~Urbach and I.~Wetzorke,
\newblock Phys. Lett. {\bf B586}, 432 (2004),
  \href{http://arxiv.org/abs/hep-lat/0312013}{{\tt hep-lat/0312013}}.

\bibitem{Jansen:2005gf}
{\bf \xlf} Collaboration, K.~Jansen, M.~Papinutto, A.~Shindler, C.~Urbach and
  I.~Wetzorke,
\newblock Phys. Lett. {\bf B619}, 184 (2005),
  \href{http://arxiv.org/abs/hep-lat/0503031}{{\tt hep-lat/0503031}}.

\bibitem{Jansen:2005kk}
{\bf \xlf} Collaboration, K.~Jansen, M.~Papinutto, A.~Shindler, C.~Urbach and
  I.~Wetzorke,
\newblock JHEP {\bf 09}, 071 (2005),
  \href{http://arxiv.org/abs/hep-lat/0507010}{{\tt hep-lat/0507010}}.

\bibitem{Abdel-Rehim:2005gz}
A.~M. Abdel-Rehim, R.~Lewis and R.~M. Woloshyn,
\newblock Phys. Rev. {\bf D71}, 094505 (2005),
  \href{http://arxiv.org/abs/hep-lat/0503007}{{\tt hep-lat/0503007}}.

\bibitem{Frezzotti:2005gi}
R.~Frezzotti, G.~Martinelli, M.~Papinutto and G.~C. Rossi,
\newblock JHEP {\bf 04}, 038 (2006),
  \href{http://arxiv.org/abs/hep-lat/0503034}{{\tt hep-lat/0503034}}.

\bibitem{Jansen:2005cg}
{\bf \xlf} Collaboration, K.~Jansen {\em et~al.},
\newblock Phys. Lett. {\bf B624}, 334 (2005),
  \href{http://arxiv.org/abs/hep-lat/0507032}{{\tt hep-lat/0507032}}.

\bibitem{Frezzotti:2007qv}
R.~Frezzotti and G.~Rossi,
\newblock PoS {\bf LAT2007}, 277 (2007),
  \href{http://arxiv.org/abs/0710.2492}{{\tt arXiv:0710.2492 [hep-lat]}}.

\bibitem{Dimopoulos:2009qv}
P.~Dimopoulos, R.~Frezzotti, C.~Michael, G.~C. Rossi and C.~Urbach,
\newblock \href{http://arxiv.org/abs/0908.0451}{{\tt arXiv:0908.0451
  [hep-lat]}}.

\bibitem{Urbach:2007rt}
{\bf ETM} Collaboration, C.~Urbach,
\newblock PoS {\bf LAT2007}, 022 (2007),
  \href{http://arxiv.org/abs/0710.1517}{{\tt arXiv:0710.1517 [hep-lat]}}.

\bibitem{Dimopoulos:2007qy}
{\bf ETM} Collaboration, P.~Dimopoulos, R.~Frezzotti, G.~Herdoiza, C.~Urbach
  and U.~Wenger,
\newblock PoS {\bf LAT2007}, 102 (2007),
  \href{http://arxiv.org/abs/0710.2498}{{\tt arXiv:0710.2498 [hep-lat]}}.

\bibitem{Dimopoulos:2008sy}
{\bf ETM} Collaboration, P.~Dimopoulos {\em et~al.},
\newblock \href{http://arxiv.org/abs/0810.2873}{{\tt arXiv:0810.2873
  [hep-lat]}}.

\bibitem{Weisz:1982zw}
P.~Weisz,
\newblock Nucl. Phys. {\bf B212}, 1 (1983).

\bibitem{Farchioni:2004ma}
F.~Farchioni {\em et~al.},
\newblock Nucl. Phys. Proc. Suppl. {\bf 140}, 240 (2005),
  \href{http://arxiv.org/abs/hep-lat/0409098}{{\tt hep-lat/0409098}}.

\bibitem{Farchioni:2004fs}
F.~Farchioni {\em et~al.},
\newblock Eur. Phys. J. {\bf C42}, 73 (2005),
  \href{http://arxiv.org/abs/hep-lat/0410031}{{\tt hep-lat/0410031}}.

\bibitem{Sommer:1993ce}
R.~Sommer,
\newblock Nucl. Phys. {\bf B411}, 839 (1994),
  \href{http://arxiv.org/abs/hep-lat/9310022}{{\tt hep-lat/9310022}}.

\bibitem{Weinberg:1978kz}
S.~Weinberg,
\newblock Physica {\bf A96}, 327 (1979).

\bibitem{Gasser:1983yg}
J.~Gasser and H.~Leutwyler,
\newblock Ann. Phys. {\bf 158}, 142 (1984).

\bibitem{Gasser:1985gg}
J.~Gasser and H.~Leutwyler,
\newblock Nucl. Phys. {\bf B250}, 465 (1985).

\bibitem{Amsler:2008zzb}
{\bf Particle Data Group} Collaboration, C.~Amsler {\em et~al.},
\newblock Phys. Lett. {\bf B667}, 1 (2008).

\bibitem{Yoshie:2008aw}
T.~Yoshie,
\newblock PoS {\bf LATTICE2008}, 019 (2008),
  \href{http://arxiv.org/abs/0812.0849}{{\tt arXiv:0812.0849 [hep-lat]}}.

\bibitem{Urbach:2005ji}
C.~Urbach, K.~Jansen, A.~Shindler and U.~Wenger,
\newblock Comput. Phys. Commun. {\bf 174}, 87 (2006),
  \href{http://arxiv.org/abs/hep-lat/0506011}{{\tt hep-lat/0506011}}.

\bibitem{Jansen:2009xp}
K.~Jansen and C.~Urbach,
\newblock Comput. Phys. Commun. {\bf 180}, 2717 (2009),
  \href{http://arxiv.org/abs/0905.3331}{{\tt arXiv:0905.3331 [hep-lat]}}.

\bibitem{Meyer:2006ty}
H.~B. Meyer {\em et~al.},
\newblock Comput. Phys. Commun. {\bf 176}, 91 (2007),
  \href{http://arxiv.org/abs/hep-lat/0606004}{{\tt arXiv:hep-lat/0606004}}.

\bibitem{Martinelli:1994ty}
G.~Martinelli, C.~Pittori, C.~T. Sachrajda, M.~Testa and A.~Vladikas,
\newblock Nucl. Phys. {\bf B445}, 81 (1995),
  \href{http://arxiv.org/abs/hep-lat/9411010}{{\tt arXiv:hep-lat/9411010}}.

\bibitem{Dimopoulos:2007fn}
P.~Dimopoulos {\em et~al.},
\newblock PoS {\bf LAT2007}, 241 (2007),
  \href{http://arxiv.org/abs/0710.0975}{{\tt arXiv:0710.0975 [hep-lat]}}.

\bibitem{Alexandrou:2009qu}
C.~Alexandrou {\em et~al.},
\newblock \href{http://arxiv.org/abs/0910.2419}{{\tt arXiv:0910.2419
  [hep-lat]}}.

\bibitem{Farchioni:2004us}
F.~Farchioni {\em et~al.},
\newblock Eur. Phys. J. {\bf C39}, 421 (2005),
  \href{http://arxiv.org/abs/hep-lat/0406039}{{\tt hep-lat/0406039}}.

\bibitem{Farchioni:2005tu}
F.~Farchioni {\em et~al.},
\newblock Phys. Lett. {\bf B624}, 324 (2005),
  \href{http://arxiv.org/abs/hep-lat/0506025}{{\tt arXiv:hep-lat/0506025}}.

\bibitem{Sharpe:2004ps}
S.~R. Sharpe and J.~M.~S. Wu,
\newblock Phys. Rev. {\bf D70}, 094029 (2004),
  \href{http://arxiv.org/abs/hep-lat/0407025}{{\tt hep-lat/0407025}}.

\bibitem{Munster:2004am}
G.~M{\"u}nster,
\newblock JHEP {\bf 09}, 035 (2004),
  \href{http://arxiv.org/abs/hep-lat/0407006}{{\tt hep-lat/0407006}}.

\bibitem{Scorzato:2004da}
L.~Scorzato,
\newblock Eur. Phys. J. {\bf C37}, 445 (2004),
  \href{http://arxiv.org/abs/hep-lat/0407023}{{\tt hep-lat/0407023}}.

\bibitem{Sharpe:2004ny}
S.~R. Sharpe and J.~M.~S. Wu,
\newblock Phys. Rev. {\bf D71}, 074501 (2005),
  \href{http://arxiv.org/abs/hep-lat/0411021}{{\tt hep-lat/0411021}}.

\bibitem{Gasser:1986vb}
J.~Gasser and H.~Leutwyler,
\newblock Phys. Lett. {\bf B184}, 83 (1987).

\bibitem{Luscher:1985dn}
M.~L{\"u}scher,
\newblock Commun. Math. Phys. {\bf 104}, 177 (1986).

\bibitem{Colangelo:2003hf}
G.~Colangelo and S.~D{\"u}rr,
\newblock Eur. Phys. J. {\bf C33}, 543 (2004),
  \href{http://arxiv.org/abs/hep-lat/0311023}{{\tt hep-lat/0311023}}.

\bibitem{Colangelo:2005gd}
G.~Colangelo, S.~D{\"u}rr and C.~Haefeli,
\newblock Nucl. Phys. {\bf B721}, 136 (2005),
  \href{http://arxiv.org/abs/hep-lat/0503014}{{\tt hep-lat/0503014}}.

\bibitem{Colangelo:2006mp}
G.~Colangelo and C.~Haefeli,
\newblock Nucl. Phys. {\bf B744}, 14 (2006),
  \href{http://arxiv.org/abs/hep-lat/0602017}{{\tt arXiv:hep-lat/0602017}}.

\bibitem{Colangelo:2001df}
G.~Colangelo, J.~Gasser and H.~Leutwyler,
\newblock Nucl. Phys. {\bf B603}, 125 (2001),
  \href{http://arxiv.org/abs/hep-ph/0103088}{{\tt arXiv:hep-ph/0103088}}.

\bibitem{Frezzotti:2008dr}
R.~Frezzotti, V.~Lubicz and S.~Simula,
\newblock Phys. Rev. {\bf D79}, 074506 (2009),
  \href{http://arxiv.org/abs/0812.4042}{{\tt arXiv:0812.4042 [hep-lat]}}.

\bibitem{Orth:2005kq}
B.~Orth, T.~Lippert and K.~Schilling,
\newblock Phys. Rev. {\bf D72}, 014503 (2005),
  \href{http://arxiv.org/abs/hep-lat/0503016}{{\tt hep-lat/0503016}}.

\bibitem{Giusti:2007hk}
L.~Giusti,
\newblock PoS. {\bf LAT2006} (2007),
  \href{http://arxiv.org/abs/hep-lat/0702014}{{\tt hep-lat/0702014}}.

\bibitem{Munster:2003ba}
G.~M{\"u}nster and C.~Schmidt,
\newblock Europhys. Lett. {\bf 66}, 652 (2004),
  \href{http://arxiv.org/abs/hep-lat/0311032}{{\tt arXiv:hep-lat/0311032}}.

\bibitem{Chetyrkin:1999pq}
K.~G. Chetyrkin and A.~Retey,
\newblock Nucl. Phys. {\bf B583}, 3 (2000),
  \href{http://arxiv.org/abs/hep-ph/9910332}{{\tt arXiv:hep-ph/9910332}}.

\bibitem{DellaMorte:2008ad}
{\bf ALPHA} Collaboration, M.~Della~Morte {\em et~al.},
\newblock JHEP {\bf 07}, 037 (2008), \href{http://arxiv.org/abs/0804.3383}{{\tt
  arXiv:0804.3383 [hep-lat]}}.

\bibitem{Durr:2008zz}
S.~D{\"u}rr {\em et~al.},
\newblock Science {\bf 322}, 1224 (2008),
  \href{http://arxiv.org/abs/0906.3599}{{\tt arXiv:0906.3599 [hep-lat]}}.

\bibitem{Necco:2009cq}
S.~Necco,
\newblock PoS {\bf CONFINEMENT8}, 024 (2008),
  \href{http://arxiv.org/abs/0901.4257}{{\tt arXiv:0901.4257 [hep-lat]}}.

\bibitem{Aoki:2009zb}
S.~Aoki,
\newblock \href{http://arxiv.org/abs/0910.3806}{{\tt arXiv:0910.3806
  [hep-lat]}}.

\bibitem{Baron:2008xa}
{\bf ETM} Collaboration, R.~Baron {\em et~al.},
\newblock PoS {\bf LATTICE2008}, 094 (2008),
  \href{http://arxiv.org/abs/0810.3807}{{\tt arXiv:0810.3807 [hep-lat]}}.

\bibitem{Baron:2009}
{\bf ETM} Collaboration, R.~Baron {\em et~al.},
\newblock PoS {\bf LATT2009}, 104 (2009).

\bibitem{Edwards:2004sx}
{\bf SciDAC} Collaboration, R.~G. Edwards and B.~Joo,
\newblock Nucl. Phys. Proc. Suppl. {\bf 140}, 832 (2005),
  \href{http://arxiv.org/abs/hep-lat/0409003}{{\tt hep-lat/0409003}}.

\bibitem{BAGEL}
P.~Boyle,
\newblock \href{http://arxiv.org/abs/http://www.ph.ed.ac.uk/\~{
  }paboyle/bagel/Bagel.html}{{\tt http://www.ph.ed.ac.uk/\~{
  }paboyle/bagel/Bagel.html}}.

\bibitem{R:2005}
{R Development Core Team},
\newblock {\em R: A language and environment for statistical computing},
\newblock R Foundation for Statistical Computing, Vienna, Austria, 2005,
\newblock {ISBN} 3-900051-07-0.

\bibitem{Lepage:2001ym}
G.~P. Lepage {\em et~al.},
\newblock Nucl. Phys. Proc. Suppl. {\bf 106}, 12 (2002),
  \href{http://arxiv.org/abs/hep-lat/0110175}{{\tt arXiv:hep-lat/0110175}}.

\bibitem{Michael:1993yj}
C.~Michael,
\newblock Phys. Rev. {\bf D49}, 2616 (1994),
  \href{http://arxiv.org/abs/hep-lat/9310026}{{\tt arXiv:hep-lat/9310026}}.

\bibitem{Michael:1994sz}
C.~Michael and A.~McKerrell,
\newblock Phys. Rev. {\bf D51}, 3745 (1995),
  \href{http://arxiv.org/abs/hep-lat/9412087}{{\tt arXiv:hep-lat/9412087}}.

\end{thebibliography}

\end{document}